\documentclass[twocolumn]{aastex63}

\newcommand{\abund}[1]{$\log N({\rm #1})/N({\rm H})$}

\newcommand{\eg}{e.g.}

\newcommand{\figref}[1]{Fig.~\ref{#1}}

\newcommand{\hone}{\ion{H}{1}}

\newcommand{\heone}{\ion{He}{1}}

\newcommand{\cone}{\ion{C}{1}}
\newcommand{\ctwo}{\ion{C}{2}}

\newcommand{\none}{\ion{N}{1}}
\newcommand{\ntwo}{\ion{N}{2}}

\newcommand{\oone}{\ion{O}{1}}

\newcommand{\neone}{\ion{Ne}{1}}

\newcommand{\mgone}{\ion{Mg}{1}}
\newcommand{\mgtwo}{\ion{Mg}{2}}

\newcommand{\altwo}{\ion{Al}{2}}

\newcommand{\sitwo}{\ion{Si}{2}}
\newcommand{\sithree}{\ion{Si}{3}}

\newcommand{\ptwo}{\ion{P}{2}}
\newcommand{\sone}{\ion{S}{1}}
\newcommand{\stwo}{\ion{S}{2}}

\newcommand{\cltwo}{\ion{Cl}{2}}

\newcommand{\catwo}{\ion{Ca}{2}}
\newcommand{\cathree}{\ion{Ca}{3}}
\newcommand{\scthree}{\ion{Sc}{3}}
\newcommand{\titwo}{\ion{Ti}{2}}
\newcommand{\tithree}{\ion{Ti}{3}}
\newcommand{\crtwo}{\ion{Cr}{2}}
\newcommand{\crthree}{\ion{Cr}{3}}
\newcommand{\mntwo}{\ion{Mn}{2}}

\newcommand{\fetwo}{\ion{Fe}{2}}

\newcommand{\cotwo}{\ion{Co}{2}}
\newcommand{\nitwo}{\ion{Ni}{2}}
\newcommand{\cutwo}{\ion{Cu}{2}}
\newcommand{\gatwo}{\ion{Ga}{2}}
\newcommand{\gathree}{\ion{Ga}{3}}

\newcommand{\pdtwo}{\ion{Pd}{2}}

\newcommand{\intwo}{\ion{In}{2}}
\newcommand{\sntwo}{\ion{Sn}{2}}
\newcommand{\hgtwo}{\ion{Hg}{2}}
\newcommand{\pbone}{\ion{Pb}{1}}
\newcommand{\pbtwo}{\ion{Pb}{2}}

\newcommand{\ebv}{$E(B-V)$\/}

\newcommand{\kms}{km s$^{-1}$}
\newcommand{\logg}{$\log g$}

\newcommand{\msun}{$M_{\sun}$}

\newcommand{\teff}{$T_{\rm eff}$}

\newcommand{\fuse}{{\em FUSE}}

\newcommand{\iue}{{\em IUE}}

\accepted{2021 June 23}
\submitjournal{AJ}

\shorttitle{The Bright Star in 47 Tuc}
\shortauthors{Dixon et al.}

\begin{document}

\title{Observations of the Bright Star in the Globular Cluster 47~Tucanae (NGC~104)}

\correspondingauthor{William V. Dixon}
\email{dixon@stsci.edu}

\author[0000-0001-9184-4716]{William V. Dixon}
\affiliation{Space Telescope Science Institute, 3700 San Martin Drive, Baltimore, MD 21218, USA}

\author[0000-0001-7653-0882]{Pierre Chayer}
\affiliation{Space Telescope Science Institute, 3700 San Martin Drive, Baltimore, MD 21218, USA}

\author[0000-0001-8031-1957]{Marcelo Miguel Miller Bertolami}
\affiliation{Instituto de Astrof\'{i}sica de La Plata, CONICET-UNLP, Paseo del Bosque s/n, FWA, B1900 La Plata, Provincia de Buenos Aires, Argentina}
\affiliation{Facultad de Ciencias Astron\'{o}micas y Geof\'{i}sicas, Universidad Nacional de La Plata, Paseo del Bosque s/n, FWA, B1900 La Plata, Provincia de Buenos Aires, Argentina}

\author[0000-0002-5176-2924]{Valentina Sosa Fiscella}
\affiliation{Facultad de Ciencias Astron\'{o}micas y Geof\'{i}sicas, Universidad Nacional de La Plata, Paseo del Bosque s/n, FWA, B1900 La Plata, Provincia de Buenos Aires, Argentina}
\affiliation{CCT La Plata, CONICET, Calle 8 N\textsuperscript{o} 1467, B1904 La Plata, Provincia de Buenos Aires, Argentina}

\author[0000-0002-8109-2642]{Robert A. Benjamin}
\affiliation{Department of Physics, University of Wisconsin-Whitewater, 800 West Main Street, Whitewater, WI 53190, USA}

\author[0000-0002-8985-8489]{Andrea Dupree}
\affiliation{Center for Astrophysics \textbar\ Harvard \& Smithsonian, 60 Garden Street, Cambridge, MA 02138, USA}



\begin{abstract}

The Bright Star in the globular cluster 47 Tucanae (NGC 104) is a post-AGB star of spectral type B8~III.
The ultraviolet spectra of late-B stars exhibit a myriad of
absorption features, many due to species unobservable from the ground.  The Bright Star
thus represents a unique window into the chemistry of 47~Tuc.  We have analyzed observations obtained with the
{\em Far Ultraviolet Spectroscopic Explorer (FUSE)}, the Cosmic Origins Spectrograph
(COS) aboard the {\em Hubble Space Telescope}, and the MIKE Spectrograph on the
Magellan Telescope.  By fitting these data with synthetic spectra, we 
determine various stellar parameters ($T_{\rm eff} = 10,850 \pm 250$ K, $\log g = 2.20 \pm 0.13$)
and the photospheric abundances of 26 elements, including 
Ne, P, Cl, Ga, Pd, In, Sn, Hg, and Pb, which have not previously been published for this cluster.  
Abundances of intermediate-mass elements
(Mg through Ga) generally scale with Fe, while the heaviest elements
(Pd through Pb) have roughly solar abundances.  Its low C/O ratio indicates that the star did not undergo third dredge-up and suggests that its heavy elements were made by a previous generation of stars.  If so,  
this pattern should be present throughout the cluster, not just in this star.
Stellar-evolution models suggest that the Bright Star is powered by a He-burning shell, 
having left the AGB during or immediately after a thermal pulse.
Its mass ($0.54 \pm 0.16 M_{\odot}$) implies that single stars in 47~Tuc lose 0.1--0.2 $M_\odot$ on the AGB, 
only slightly less than they lose on the RGB.

\end{abstract}




\section{Introduction} \label{sec:intro}

The Bright Star in the globular cluster 47~Tucanae (NGC~104) is the cluster's brightest member at both ultraviolet and optical wavelengths.  It is Star No. \textsc{i} in the catalog of \citet{Feast:Thackeray:60}, who refer to it as ``the bright blue star.''  It is star R1 in the catalog of \citet{LloydEvans:74}.  In his study of UV-bright stars in globular clusters, \citet{deBoer:1985} calls the star ``BS.'' 

This blue giant (B8 III) is a post-asymptotic giant branch (post-AGB) star.  Having ascended the AGB, it is moving across the color-magnitude diagram toward the tip of the white-dwarf cooling sequence.  \citet{Dixon:95} observed the star with the Hopkins Ultraviolet Telescope \citep[HUT;][]{HUT1CAL1}.  By comparing the shape of its far-ultraviolet spectrum with those predicted by \citet{Kurucz:92} models, they derived an effective temperature \teff\ = 11,000 K and a surface gravity \logg\ = 2.5.

While Kurucz models with [M/H] = $-1.0$ adequately reproduce the stellar spectrum across most of the HUT bandpass (900--1850 \AA), they over-predict the flux at wavelengths shorter than Lyman $\beta$ (1026 \AA).  To understand this discrepancy, \citet{Dixon:Chayer:2013} observed the star with the {\em Far Ultraviolet Spectroscopic Explorer (FUSE)}.  They determined that the peculiar absorption features in the stellar continuum are sculpted by resonances in the photoionization cross section of excited-state neutral nitrogen in the stellar atmosphere.

\begin{deluxetable*}{lcccclcc}
\tablecaption{Summary of {\it FUSE}, COS, and Magellan Observations \label{tab:log_obs}}
\tablehead{
\colhead{Instrument} & \colhead{Grating} & \colhead{Wavelength} & \colhead{$R\equiv\lambda/\Delta\lambda$} & \colhead{Exp. Time} & \colhead{Obs.\ Date} & \colhead{Data ID} & \colhead{P.I.} \\
& & \colhead{(\AA)} & & \colhead{(s)}
}
\startdata
{\it FUSE} & $\cdots$ &\phn905--1187 & 20,000 &   14,058 & 2006 Jun 7 & G0730101 & Dixon \\
COS & G130M & 1130--1430 & 18,000 & \phn\phn\phd524 & 2015 Mar 26 & LCKV01010 & Benjamin \\
COS & G160M & 1410--1780 & 18,000 & \phn\phn\phd764 & 2015 Mar 26 & LCKV01020 & Benjamin \\
MIKE-Blue & $\cdots$ & 3350--5000 & 36,000 & \phn\phn\phd900 & 2010 Jun 30 & Obj\_mb0063 & Dupree \\
MIKE-Red & $\cdots$ & 4900--9300 & 26,000 & \phn\phn\phd900 & 2010 Jun 30 & Obj\_mr0057 & Dupree \\
\enddata
\tablecomments{The MAST data used in this work are available at \dataset[10.17909/t9-6s10-5z38]{https://doi.org/10.17909/t9-6s10-5z38}.}
\end{deluxetable*}

None of these studies attempted to derive the Bright Star's chemical abundances.  The spectra of late-B stars exhibit a myriad of metal-line absorption features.  The Bright Star thus represents a unique window into the chemistry of 47~Tuc.  To take advantage of this opportunity, we have combined the star's \fuse\/ spectrum with data obtained with the Cosmic Origins Spectrograph (COS) aboard the {\em Hubble Space Telescope}\/ and the MIKE Spectrograph on the Magellan Telescope.  A summary of these observations is presented in Table~\ref{tab:log_obs}.

In Section \ref{sec_observations}, we discuss each of our data sets.  In Section \ref{sec_analysis}, we present our atmospheric models and use them to derive stellar parameters and abundances.  In Section \ref{sec_models_discussion}, we consider various problems with the models, some that we are able to solve, and some that we are not.  In Section \ref{sec_discussion}, we discuss our results: 
we address the question of cluster membership in Section \ref{sec_membership},
we derive the stellar mass and luminosity in Section \ref{sec_mass},
we  consider the evolutionary state of the star in Section \ref{sec_evolution}, and
we consider the implications of our chemical abundances in Section \ref{sec_abundance_discussion}.
We summarize our conclusions in Section \ref{sec_conclusions}.

\section{Observations and Data Reduction}\label{sec_observations}

\subsection{{\em FUSE}\/ Spectroscopy}

\fuse\/ provides medium-resolution spectroscopy from 1187 \AA\ to the Lyman limit \citep{Moos:00, Sahnow:00}.  The Bright Star was observed through the \fuse\/ $30\arcsec \times 30\arcsec$ aperture.  The data were reduced using v3.2.2 of CalFUSE, the standard data-reduction pipeline software \citep{Dixon:07}, and retrieved from the Mikulski Archive for Space Telescopes (MAST).  The signal-to-noise ratio (S/N) per resolution element is $< 5$ at wavelengths shorter than 1026 \AA\ and 10--15 at longer wavelengths.

\subsection{COS Spectroscopy}

The Cosmic Origins Spectrograph (COS) enables high-sensitivity, medium- and low-resolution spectroscopy in the 1150--3200 \AA\ wavelength range \citep{Green:COS:2012}.  The Bright Star was observed with COS using both the G130M and G160M gratings. The fully-reduced spectra, processed with CALCOS version 3.3.7 \citep{Rafelski:2018}, were retrieved from MAST.  The S/N per resolution element varies between 20 and 40 in the G130M spectrum.  In the G160M spectrum, the S/N falls monotonically from $\sim 55$ at 1430 \AA\ to $\sim 20$ at 1770~\AA.

In the course of our analysis, we will have some questions about the interstellar medium (ISM) along the line of sight to the Bright Star.  Fortunately, the cluster star UIT~14 \citep{OConnell:1997} has been observed with both the G130M and G160M gratings of COS (Program 11592; P.I. Lehner).  The star is only 1\arcmin\ from our target, so samples the same line of sight.  Its high effective temperature (\teff\ $\sim 50,000$ K) ensures that all low-ionization features in its spectrum are interstellar.  

\subsection{MIKE Spectroscopy}\label{sec_MIKE}

The Bright Star was observed with the Magellan/Clay telescope at the Las Campanas Observatory using the Magellan Inamori Kyocera Echelle (MIKE) spectrograph \citep{Bernstein:2003} with a $0.70 \times 5$ arcsecond slit.  The data were obtained as part of the observing run documented in \citet{Dupree:2011} and were reduced using the procedures described therein.  The authors quote a spectral resolution of $\sim$ 26,000 on the blue side and $\sim$ 36,000 on the red, but these values are reversed; the resolution is higher on the blue side.  The S/N per resolution element varies within each spectral order and with wavelength, reaching a maximum of about 450 at 4700 \AA\ on the blue side and 7500 \AA\ on the red.  The spectrum extends from the Balmer edge to the Paschen series (lines P10 and higher) of hydrogen.

The orders of an echelle spectrum have a distinctive shape, which we must remove before attempting to fit individual features.  \citet{Dupree:2011}\ normalized their spectra with a cubic spline fit to the spectrum of the standard white dwarf LTT~9491, but this curve does not perfectly reproduce the shape of the Bright Star spectrum.  In scaling the curve to match our data, we have optimized the fit in regions around the hydrogen Balmer lines.  Elsewhere, particularly near the edges of an order, a small, residual tilt is often apparent in the normalized spectrum.

\subsection{Contamination}\label{sec_contamination}

Globular clusters are crowded fields, and the Bright Star lies within an arc minute of the cluster center.  While the COS aperture is only 2\farcs5 in diameter, the \fuse\/ aperture is much larger.   Fortunately, most bright cluster stars are red giants and invisible in the ultraviolet.  GALEX FUV images of the cluster reveal no bright objects within 20\arcsec\ of our target, and the cluster's FUV vs.\ (FUV--NUV) color-magnitude diagram shows that the Bright Star is nearly three magnitudes brighter than any other star in the field \citep{Schiavon:2012}.  In the optical, contamination is a larger threat.  The MIKE aperture was deliberately positioned to exclude two bright, nearby stars.  (See the identification chart in \citealt{Forte:2002}.)  As discussed in Section \ref{sec_abundance}, the fact that we derive the same high nitrogen abundance  from both FUV and near-IR features suggests that contamination of the MIKE spectrum is insignificant.

\section{Spectral Analysis} \label{sec_analysis}

\subsection{Model Atmospheres}\label{sec_models}

Because 47~Tuc has a relatively high metallicity ([Fe/H] = $-0.74$; \citealt{Kovalev:2019}), and because the spectra of B-type stars exhibit many absorption features, we expect line opacity to play an important role in the atmospheric structure of the Bright Star.  We therefore employ the model atmospheres of \citet{Castelli:Kurucz:03}.  These models provide a thorough treatment of line blanketing by metals, though the assumption of local thermodynamic equilibrium (LTE) neglects non-LTE effects, which can be important for hot, low-gravity stars.  We use the $\alpha$-enhanced models ([$\alpha$/Fe] = +0.4) that are generally appropriate for stars in a globular cluster.  We use models with metallicity [Fe/H] = $-1.0$, microturbulent velocity $\xi$ = 2.0 \kms, and mixing-length parameter $l/H$ = 1.25.  From these model atmospheres, we compute synthetic spectra using version 51 of the program SYNSPEC \citep{Hubeny:88}.  Our atomic models are similar to those used by \citet{Lanz:Hubeny:2007} to compute their grid of B-type stars.  

Because we use pre-computed model atmospheres, we cannot change their chemical abundances.  Instead, we adjust the abundances used by SYNSPEC to calculate the model spectra.  This technique makes the implicit assumption that small changes in the photospheric abundances do not significantly alter the star's atmospheric structure.  While this is certainly true for most metals, it is probably not the case for helium, as discussed below.

Before comparison with the data, synthetic spectra must be convolved with the appropriate line-spread function.  For the \fuse\/ spectrum, we use a Gaussian of FWHM = 0.06 \AA.  For the COS spectra, we employ the tabulated line-spread functions appropriate for data obtained at Lifetime Position \#3, which are available from \href{http://www.stsci.edu/hst/cos/performance/spectral_resolution/}{the COS website}.  For the MIKE spectra, we assume a resolving power of R = 36,000 on the blue side and 26,000 on the red, convolving the spectrum with a Gaussian of FWHM = $\lambda/$R, where $\lambda$ is the wavelength of the feature that we wish to fit.  The resulting synthetic spectra successfully reproduce the observed absorption features; in particular, we find no evidence for rotational line broadening.

\subsection{Spectral Fitting}\label{sec_fitting}

Given a grid of synthetic spectra, our fitting routine linearly interpolates among them---in one, two, or three dimensions, as appropriate---determining the best fit to the data via chi-squared minimization.  The uncertainties quoted for parameters derived from individual line fits are $1 \sigma$ errors computed from the covariance matrix returned by the fitting routine; we refer to these as statistical errors.   In most cases, we ignore statistical errors when combining the results from multiple line fits.  Instead, we simply quote the mean and standard deviation of the various measurements.  We believe that this approach is more appropriate when systematic uncertainties---such as the shape and level of the stellar continuum---are the dominant source of error.

\subsection{Stellar Parameters}\label{sec_parameters}

We begin by constructing a three-dimensional grid of synthetic spectra with effective temperatures \teff\ = 10,000 to 11,750 K in steps of 250 K, surface gravities \logg\ = 2.0 and 2.5, and helium abundances \abund{He} = $-0.6$ to $-1.4$ in steps of 0.2.  Models with higher temperatures or lower gravities are not available.  The Castelli-Kurucz models assume a helium abundance of 10\% by number.  We compute synthetic spectra with other He/H ratios by adjusting the atmospheric abundances within SYNSPEC.

\begin{figure}
\plotone{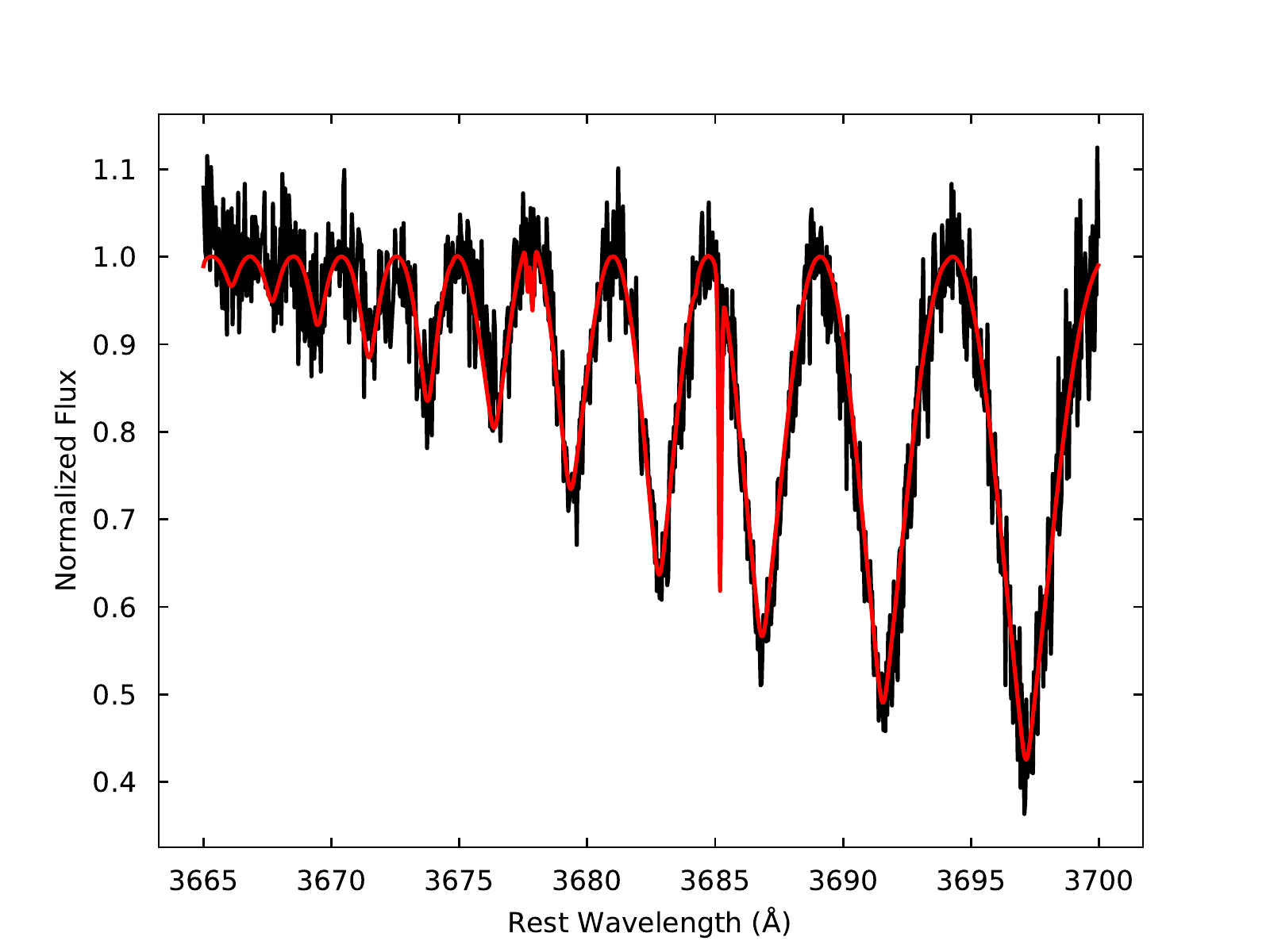}
\caption{High-order Balmer lines in the Bright Star's optical spectrum.  The black curve represents the data, the red curve our best-fit model.}
\label{fig_edge}
\end{figure}

\begin{figure*}
\plottwo{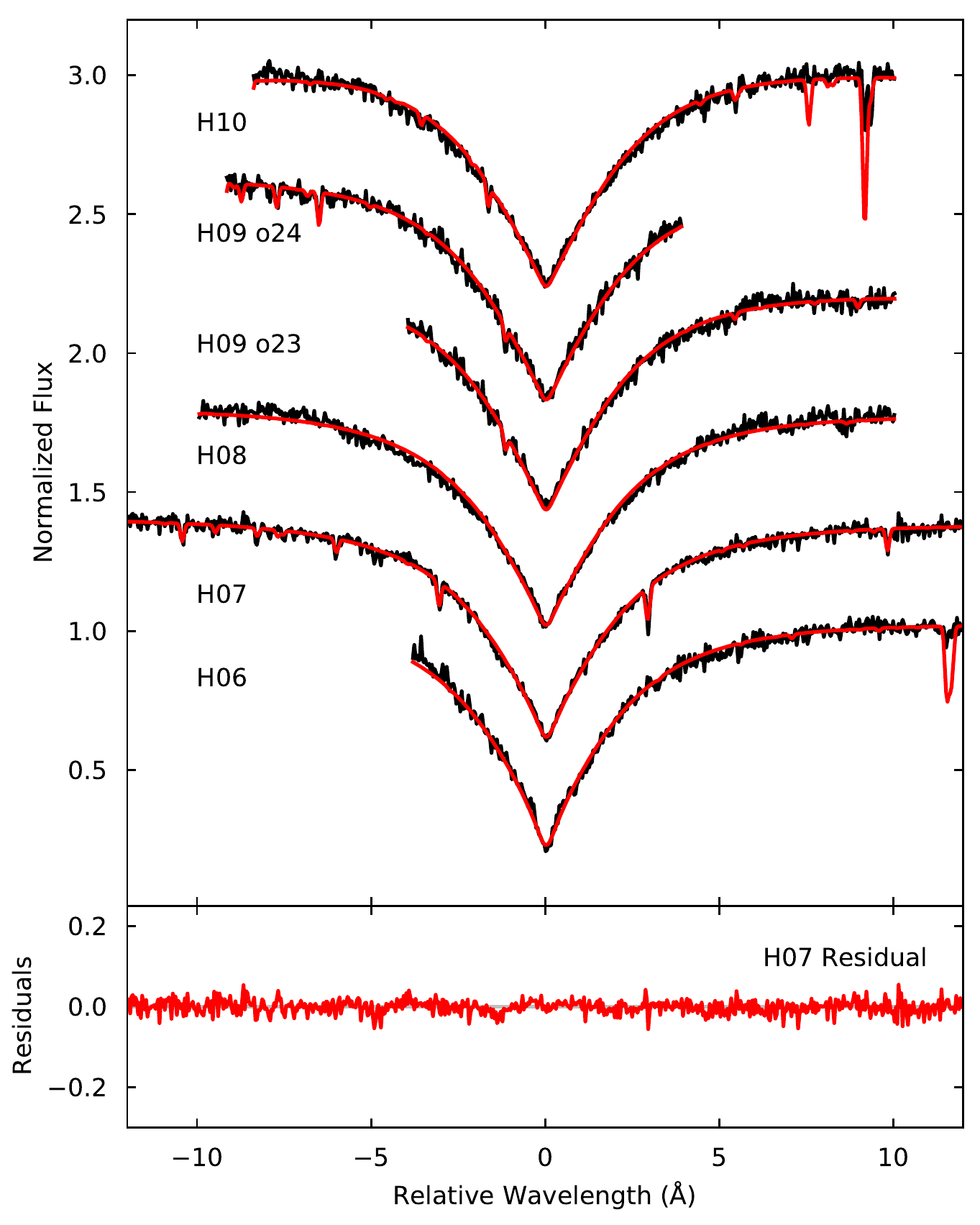}{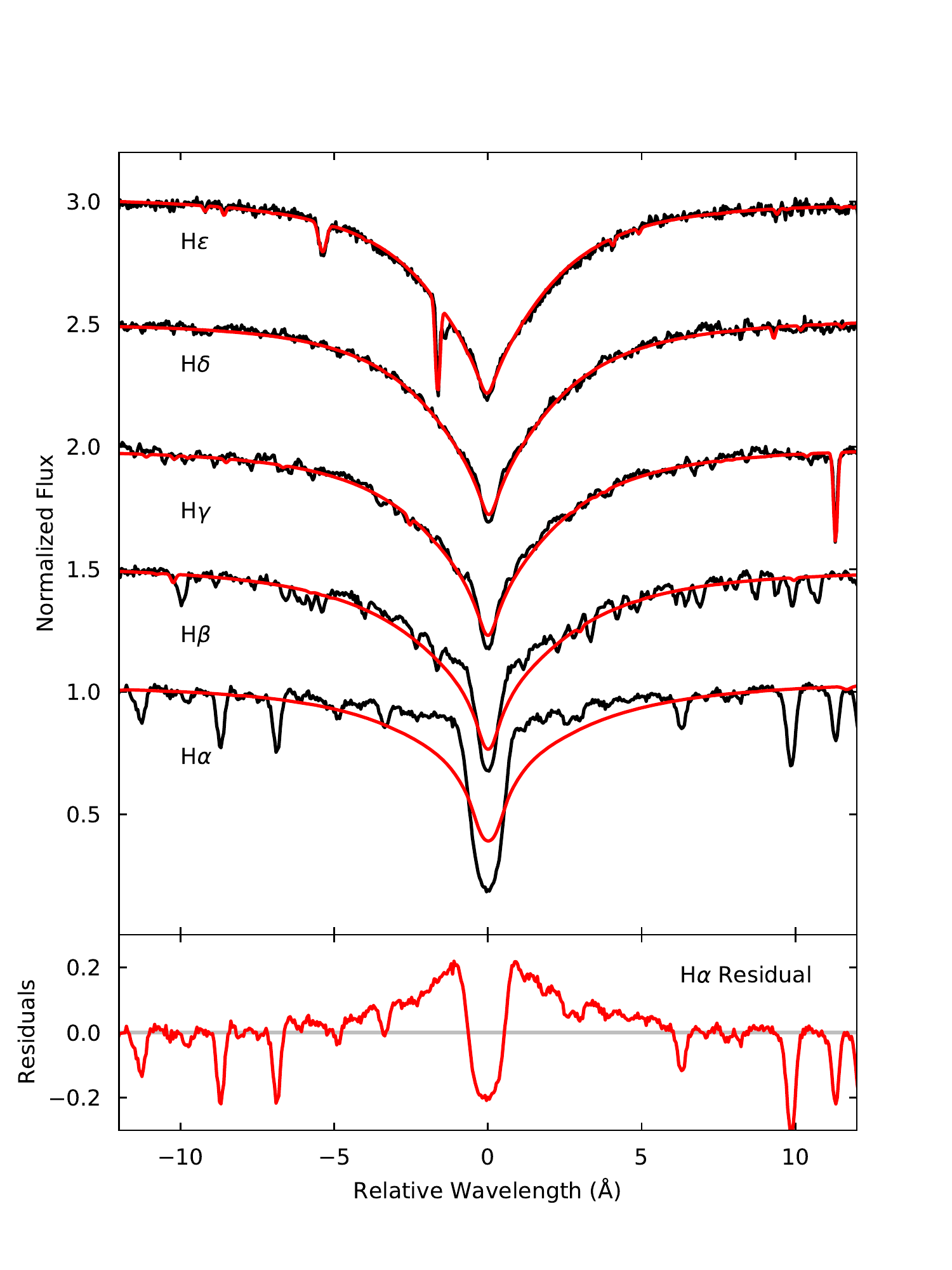}
\caption{Low-order Balmer lines in the Bright Star's optical spectrum.  Black curves represent the data, red curves our best-fit model.}
\label{fig_balmer}
\end{figure*}

Numerical experiments show that the shape of the hydrogen series near the Balmer jump is exquisitely sensitive to the surface gravity, while the low-order Balmer features provide a better probe of effective temperature. We thus determine these parameters using an iterative technique: (1) fix the temperature and fit the high-order lines (3665--3700 \AA) to determine the gravity; (2) fix the gravity and fit the low-order lines to determine the temperature (in this step, we consider only the H6--H10 lines, as discussed below); (3) repeat until the values converge.  Our best-fit parameters are \teff\ = 10,850 K and \logg\ = 2.20.  For these fits, we fix the helium abundance at \abund{He} = $-1.0$.

Our results are presented in Figs.\ \ref{fig_edge} and \ref{fig_balmer}.  We see that the high-order and H6--H10 lines are well fit by our model spectrum.  We are less successful with the H$\alpha$--H$\epsilon$ lines: the model features are too wide and too shallow.  The H$\alpha$ line is particularly troublesome, because the fitting routine drives the best-fit model beyond the limits of our grid.  We will return to this topic in Section \ref{sec_models_discussion}; for now, we consider only the H6--H10 lines. 

To estimate the uncertainty in these parameters, we determine them again in a slightly different way.  We re-fit each of the H6--H10 lines, this time allowing both the effective temperature and surface gravity to vary.  The resulting parameters are \teff\ = 11,100 K and \logg\ = 2.33, where these values are averages from fits to each of our six lines.  (Six, because the H09 line appears in two orders, and we fit each copy separately.)  We take the difference between these and our previous results to be the uncertainty in each parameter.  Our final parameters are thus \teff\ = $10,850 \pm 250$ K and \logg\ = $2.20 \pm 0.13$.

Before fitting the high-order lines, we normalize the appropriate echelle order (\#26 in the blue spectrum) by hand, identifying the peaks between each absorption line, fitting a spline to them, and dividing by the spline.  We use the same technique to normalize the model spectra.  

The echelle orders containing the low-order lines are normalized as discussed in Section \ref{sec_MIKE}.  A residual tilt is apparent in the some regions of the normalized spectra.  To account for this tilt, we multiply our model spectra by a linear function whose slope and y-intercept are free parameters in the fit.  We mask all nearby metal lines to prevent them from affecting the continuum fit.

\subsection{Helium Abundance}\label{sec_helium}

\begin{figure}
\plotone{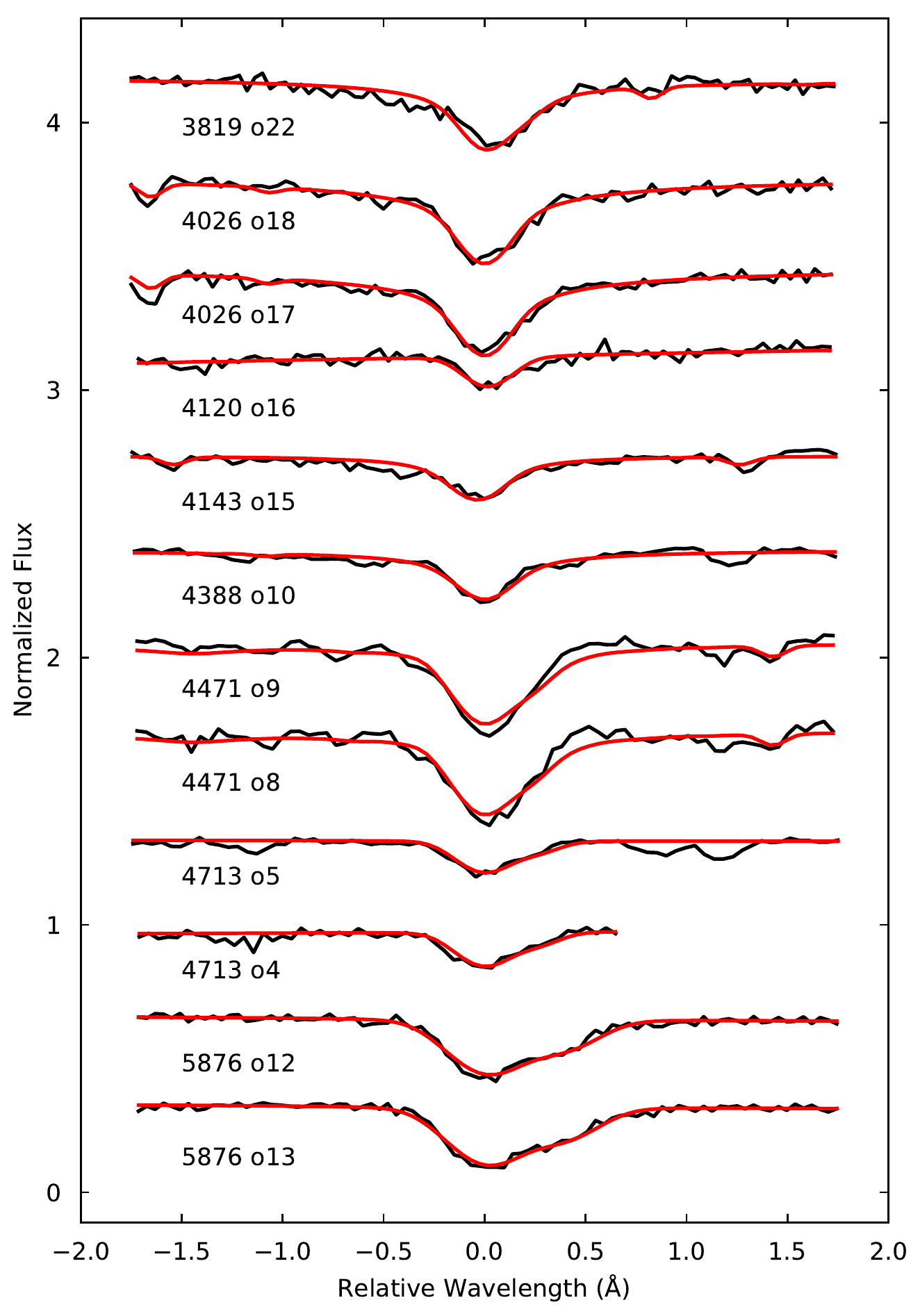}
\caption{Helium lines in the Bright Star's optical spectrum.  Black curves represent the data, red curves our best-fit models.  Lines appearing in multiple orders are fit separately.}
\label{fig_helium}
\end{figure}

To constrain the star's helium abundance, we fit selected \heone\ features in its optical spectrum.  When features appear in multiple orders, we fit each copy separately.  Features blended with metal lines are omitted.  We use the model grid described above, but fix \teff\ and \logg\ at their best-fit values and allow only the He/H ratio to vary.  Results are presented in Table \ref{tab:helium} and \figref{fig_helium}.  We derive a helium abundance \abund{He} = $-0.82 \pm 0.16$, where the values represent the mean and standard deviation of our twelve measurements.

This value of the helium abundance corresponds to a mass fraction $Y = 0.37 \pm 0.08$, which is considerably greater than the range $Y = 0.26$--0.29 derived for the various subpopulations of 47~Tuc \citep{Heyl:2015, Salaris:2016, Denissenkov:2017}.  This apparent helium enhancement may be a non-LTE effect. \citet{Takeda:2000} determined the helium abundances of a sample of late-B and early-A type supergiants.  For the four stars whose effective temperature and surface gravity are closest to those of the Bright Star, they found that LTE models over-predict the helium abundance by 0.57--0.79 dex relative to non-LTE models. 

Another possible explanation reflects a limitation in our models:  Because our synthetic spectra are derived from atmospheric models with a fixed helium abundance (10\% by number), they do not reflect variations in the atmospheric structure that would accompany changes in the He/H ratio.  Indeed, synthetic spectra computed assuming  \abund{He} = $-0.82$ yield the same values of the mean N and Fe abundance as spectra computed assuming  \abund{He} = $-1.0$.

\begin{deluxetable}{cccc}
\tablecaption{Helium Features \label{tab:helium}}
\tablehead{
\colhead{$\lambda_{\rm lab}$} & \colhead{Spectrum} & \colhead{Order} & \colhead{\abund{He}}
}
\startdata
 3819.6 & Blue & 22 &   $-0.86 \pm 0.04$ \\
 4026.2 & Blue & 18 &   $-0.62 \pm 0.03$ \\
 4026.2 & Blue & 17 &   $-0.61 \pm 0.03$ \\
 4120.8 & Blue & 16 &   $-1.12 \pm 0.06$ \\
 4143.8 & Blue & 15 &   $-0.78 \pm 0.02$ \\
 4387.9 & Blue & 10 &   $-0.72 \pm 0.02$ \\
 4471.5 & Blue & \phn9 &        $-0.71 \pm 0.02$ \\
 4471.5 & Blue & \phn8 &        $-0.68 \pm 0.03$ \\
 4713.1 & Blue & \phn5 &        $-1.04 \pm 0.03$ \\
 4713.1 & Blue & \phn4 &        $-0.96 \pm 0.05$ \\
 5875.6 & Red & 12 &   $-0.91 \pm 0.03$ \\
 5875.6 & Red & 13 &   $-0.85 \pm 0.02$ \\
\enddata
\end{deluxetable}

\subsection{Metal Abundances}\label{sec_abundance}

We derive the star's metal abundances by fitting synthetic spectra to selected absorption features in its observed spectrum.  For each element, we use the line list produced by SYNSPEC to identify the strongest features in the bandpass of each instrument.  We omit features that are likely to be blended with geocoronal emission or photospheric, interstellar, or telluric absorption lines.  A few exceptions are described below.  Our final list of absorption features is presented in Tables~\ref{tab:lines_fuv} and \ref{tab:lines_mike}, found in the Appendix.

Because we lack a model atmosphere with our best-fit stellar parameters, we take the four Castelli-Kurucz models that are closest in parameter space (\teff = 10,750 K and 11,000 K; \logg = 2.0 and 2.5), generate synthetic spectra for each, and interpolate among them to our best-fit values.  We adjust the photospheric abundances within SYNSPEC.  We repeat the process to generate a grid of spectra with a range of abundance values for the element of interest.

At FUV wavelengths, our synthetic spectra suffer a serious flaw.  As discussed by \citet{Dixon:Chayer:2013}, the Bright Star's spectrum exhibits broad absorption features due to resonances in the photoionization cross section of excited-state neutral nitrogen.  The \citet{Lanz:Hubeny:2007}  atomic models include these resonances, but smoothed over many tens of \AA ngstroms.  The resulting synthetic spectra exhibit strange, sloping continua.  Fortunately, they are easily repaired:  For each synthetic spectrum, SYNSPEC writes a continuum model to a separate file.  We divide by this continuum, then multiply by the continuum of a spectrum generated with no metals---and thus no resonant absorption.

Continuum placement is the dominant uncertainty in our fits to the \fuse\/ and COS spectra.  The FUV spectra of B-type stars are riddled with absorption lines.  Particularly in the \fuse\/ bandpass, the apparent continuum lies far below the true continuum level.  After considerable experimentation, we developed the following scheme to estimate the FUV continuum.  We scale the synthetic spectrum by a \citet{Fitzpatrick:1999} extinction curve assuming \ebv\ = 0.04 and an ISM absorption model assuming an \hone\ column density log $N$(\hone) = 21.90 [cm$^{-2}$].  (Our choice of \hone\ column density is discussed in Section \ref{sec_lyman_discussion}.)  Finally, we scale the model to reproduce the continuum in the region around 1500 \AA, where the S/N is highest.

As discussed below, we determine the iron abundance from the star's FUV spectrum by dividing it into 10 \AA\ pieces, masking the lines not due to iron, and fitting model spectra to each piece.  Considering only the well-behaved long-wavelength segment of the G160M spectrum, we derive an iron abundance that is consistent with that derived from the optical.  By inverting this process, we can test the accuracy of our continuum model.  Continuing to shorter wavelengths, we determine the factor by which our model spectrum must be scaled to yield the optically-determined iron abundance.  We find that this scale factor is within a few percent of unity down to about 1180 \AA, at which point our model continuum becomes less reliable.  When possible, we avoid using lines with wavelengths shorter than 1180 \AA.  When we must use such lines, we multiply our models by a scale factor determined from nearby iron features.  

We estimate an uncertainty of 5\% in our FUV continuum placement.  To see how this uncertainty propagates into our abundance estimates, we perform each fit twice, once with the model spectrum scaled as described and again with the model scaled by an additional factor of 0.95.  The difference in the two abundances is an estimate of the systematic error due to uncertainties in the continuum level. We add this term and the statistical error in quadrature to compute the uncertainty reported for each feature.  At optical wavelengths, we use only the statistical error.

\begin{figure*}
\plotone{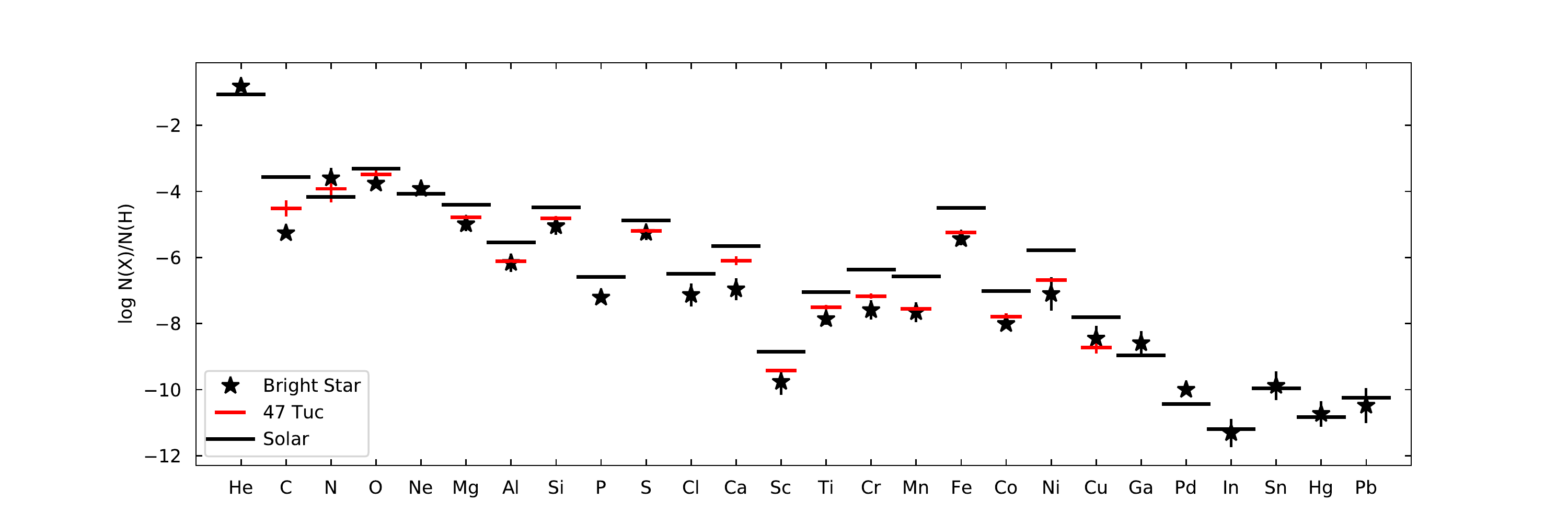}
\caption{Photospheric abundances of the Bright Star (stars), 47~Tuc (red lines), and the solar photosphere (black lines).  See  Table \ref{tab:abundance} for references.}
\label{fig:abundance}
\end{figure*}

We fit each line or group of lines separately.   We measure the equivalent width for the optical lines, but not for the FUV features, which are often blended and for which the continuum level is uncertain.  Because the scale factors of both the data and the synthetic spectra are fixed, the only free parameter in our fits is the abundance of the element in question.  For each element, the final abundance and its uncertainly represent the mean and the standard deviation of our individual measurements.  Results are presented in Table \ref{tab:abundance} and plotted in \figref{fig:abundance}.  Notes on individual elements follow.

{\em Carbon:}  The Bright Star's FUV spectrum exhibits absorption from both \cone\ and \ctwo.  The \cone\ lines yield an abundance \abund{C} = $-6.16 \pm 0.15$, a value more than 1.5 dex below the cluster mean.  The star's lone \ctwo\ line yields \abund{C} = $-5.26 \pm 0.13$.  \citet{Alexeeva:2016} compute non-LTE models of stars with effective temperatures ranging from 9550 to 17,500 K, though with surface gravities and metallicities higher than those of the Bright Star.  In each of their models, \ctwo\ dominates the total carbon abundance throughout the line-formation region of the atmosphere. The fraction of neutral carbon does not exceed a few tenths of a percent, and in the line-formation layers, it is smaller in non-LTE than in LTE models due to ultraviolet (UV) overionization.  We therefore adopt the carbon abundance derived from the \ctwo\ line as the stellar value.

{\em Nitrogen:}\/  The Bright Star's spectrum exhibits numerous \none\ features.  The optical lines yield \abund{N} = $-3.48 \pm 0.11$, and the FUV lines yield \abund{N} = $-3.78 \pm 0.40$.  The spectrum contains three strong \ntwo\ lines, one in the FUV and two in the optical; they yield \abund{N} = $-3.44 \pm 0.01$.   As these values are consistent within the uncertainties, we use the mean of all 20 lines as the stellar value.

\citet{Przybilla:Butler:2001} compare LTE and non-LTE models for several late-B and early-A type supergiants.  For the two stars whose parameters are most similar to the Bright Star, they find that LTE models over-predict the nitrogen abundance.  For $\eta$ Leo (\teff\ = 9600 K, \logg\ = 2.00), the \none\ lines modeled in LTE overestimate the abundance by 0.36 dex on average and the \ntwo\ lines by 0.15 dex.  For $\beta$ Ori (\teff\ = 12,000 K, \logg\ = 1.75), the corresponding values are 0.58 dex for \none\ and 0.27 dex for \ntwo.  Thus, our nitrogen abundance may be slightly overestimated.

As discussed in Section \ref{sec_contamination}, the Bright Star lies near other, redder stars with which its spectrum may be blended at long wavelengths.  To determine whether our MIKE spectrum is contaminated, we consider the nitrogen abundance derived from lines at the longest wavelengths.  Nitrogen is useful for this test, because its abundance is greatly enhanced relative to the cluster mean.  All of the optical \none\ features have wavelengths longer than 7400 \AA.  As we have seen, they yield an abundance consistent with that of the FUV features, indicating that contamination of the MIKE spectrum is insignificant.  

To fit the nitrogen lines with wavelengths between 1165 and 1178 \AA, we rescale our model spectra by a factor of 0.93, a value determined from fits to nearby iron lines. 

{\em Oxygen:}\/ We adopt the mean abundance derived from all of our \oone\ lines as the stellar value.  \citet{Franchini:2021} point out that, among the optical \oone\ features, the high-excitation 6158 \AA\ line is less susceptible to non-LTE effects than the \oone\ triplet at 7772, 7774, and 7775 \AA.  The 6158 \AA\ line yields an abundance \abund{O} = $-3.97 \pm 0.03$, slightly lower than our adopted value.  

To fit the \oone\ $\lambda 1152$ features, we rescale our model spectra by a factor of 0.86, again determined from fits to nearby iron lines.

{\em Aluminum:}\/  The strongest aluminum feature in the Bright Star's FUV spectrum is the \altwo\ line at 1670 \AA.  Being a resonance transition, it is also a prominent ISM feature.  We fit a Gaussian profile to the ISM line in the G160M spectrum of UIT~14.  Including this ISM line in our fit to the \altwo\ $\lambda 1670$ line in the Bright Star spectrum yields \abund{Al} $= -6.20 \pm 0.09 $. Ignoring the ISM contribution raises the derived abundance to $-6.17$, a change smaller than the uncertainty in our fit.  We adopt the mean value from all  \altwo\ features as the stellar abundance.

{\em Phosphorus:}\/  To fit the \ptwo\ features near 1150 \AA, we rescale our model spectra by a factor of 0.87.

{\em Sulfur:}\/  All of the available sulfur lines in the star's FUV spectrum are due to \sone; they yield a mean abundance \abund{S} $= -6.11 \pm 0.14$, a full dex lower than the cluster mean.  The optical lines are due to \stwo\ and yield a mean abundance \abund{S} $= -5.25 \pm 0.22$.  We attribute this discrepancy to non-LTE effects. At the effective temperature of the Bright Star, \stwo\ is the dominant species; we thus adopt the optically-derived abundance for this element.

{\em Chlorine:}\/  Our model spectra predict numerous chlorine features in the star's FUV spectrum, but all are hopelessly blended with other photospheric lines except for a pair of \cltwo\ lines at 1075 and 1079 \AA.  Of these, the 1075 \AA\ feature falls in a region where the continuum is depressed by a broad bound-free transition of \none\ (Fig.~3 of \citealt{Dixon:Chayer:2013}) that is not included in our model.  Our chlorine abundance is thus derived from a single \cltwo\ line at 1079 \AA.  For this feature, we rescale our model spectra by a factor of 1.13, a value determined from fits to nearby iron lines.

{\em Calcium:}\/  Only the \catwo\ $\lambda 1288$ feature yields an abundance consistent with the cluster mean.  The line falls at the edge of a COS detector segment, where the S/N ratio is low and the wavelength scale uncertain.  The other two FUV lines and all three of the optical features, also due to \catwo, yield consistent and much lower abundances.  Again, non-LTE effects may be responsible.  At the effective temperature of the Bright Star, \cathree\ is the dominant species \citep{Sitnova:2018}.  Unfortunately, there are no \cathree\ features in our spectra, so we quote the mean of our \catwo\ results.

{\em Titanium:}\/  The two ionization states of titanium yield different abundances.  The optical lines are due to \titwo\ and yield \abund{Ti} $= -8.43 \pm 0.19$, while the FUV lines are due to \tithree\ and yield \abund{Ti} $= -7.86 \pm 0.18$.  The cluster abundance is \abund{Ti} = $-7.50 \pm 0.07$ \citep{Kovalev:2019}.  \citet{Sitnova:2016} compute non-LTE models of stars with \teff\ between 5780 and 12800 K.  They find that non-LTE effects for \titwo\ are small for the coolest models, but grow toward higher \teff\ and lower \logg\ and [Fe/H].  In their hottest model, \tithree\ is the dominant species.  We adopt the \tithree\ abundance for this element.

{\em Iron:}\/  We fit 17 individual \fetwo\ lines in the MIKE spectrum; they yield a mean abundance \abund{Fe} = $-5.46 \pm 0.18$.  There are hundreds of iron lines in the FUV spectrum.  Instead of fitting individual features, we divide the long-wavelength segment of the G160M spectrum into 10 \AA\ pieces.  We mask ISM features and all stellar features that are not due to iron and fit our model spectra to each piece.  We omit a few pieces that contain broad absorption features or lack unblended iron lines.  The 14 remaining pieces, spanning the wavelength range 1610--1760 \AA, yield a mean iron abundance \abund{Fe} = $-5.43 \pm 0.19$, consistent with our optically-determined value.

{\em Nickel:}\/  The large error bar on our nickel abundance, 0.51 dex, reflects the wide range of values derived from individual features.  Several features, \nitwo\ $\lambda 1501, \lambda 1502, \lambda 1536.9,$ and $\lambda 3849$, yield values consistent with the cluster mean, but the rest are much lower.  We adopt the mean of all available measurements as the stellar value.

{\em Gallium:}\/  \citet{Nielsen:2005} fit LTE models to the \gatwo\ and \gathree\ features in a high-resolution STIS spectrum of the star $\chi$ Lup (\teff\ = 10,650 K, \logg\ = 3.8).  They found that the ground-state resonance lines \gatwo\ $\lambda 1414$, \gathree\ $\lambda 1495$, and \gathree\ $\lambda 1534$ yield abundances much greater than the lines from excited states of \gatwo.  In our data, \gatwo\ $\lambda 1414$ yields an abundance almost 2.0 dex less than the excited-state transitions, while \gathree\ $\lambda 1495$ and $\lambda 1534$ yield abundances consistent with them.  \cite{Castelli:2017} report that \gatwo\ is the dominant species in the photosphere of HR~6000 (\teff\ = 13,450 K, \logg\ = 4.3), so we assume that the same is true for the Bright Star.  We thus use only the excited-state transitions of \gatwo\ to derive the abundance of this species.  Note: Our original line list included only the \gatwo\ $\lambda 1414$ feature.  We employ the line parameters published by \citeauthor{Nielsen:2005} to model the gallium features in our spectrum.

\begin{deluxetable}{lccc}
\caption{Photospheric Abundances \label{tab:abundance}}
\tablehead{
\colhead{Species} & \colhead{Bright Star} & \colhead{47~Tuc} & \colhead{Sun}
}
\startdata
He & $-0.82 \pm 0.16$ & \nodata & $-1.07 \pm 0.01$ \\
C  & $-5.26 \pm 0.13$ & $-4.51 \pm 0.24$ & $-3.57 \pm 0.05$ \\
N  & $-3.60 \pm 0.31$ & $-3.92 \pm 0.41$ & $-4.17 \pm 0.05$ \\
O  & $-3.76 \pm 0.15$ & $-3.49 \pm 0.18$ & $-3.31 \pm 0.05$ \\
Ne & $-3.92 \pm 0.08$ & \nodata & $-4.07 \pm 0.10$ \\
Mg & $-4.99 \pm 0.21$ & $-4.78 \pm 0.05$ & $-4.40 \pm 0.04$ \\
Al & $-6.16 \pm 0.28$ & $-6.12 \pm 0.11$ & $-5.55 \pm 0.03$ \\
Si & $-5.05 \pm 0.27$ & $-4.81 \pm 0.06$ & $-4.49 \pm 0.03$ \\
P  & $-7.21 \pm 0.10$ & \nodata & $-6.59 \pm 0.03$ \\
S  & $-5.25 \pm 0.22$ & $-5.19 \pm 0.14$ & $-4.88 \pm 0.03$ \\
Cl & $-7.13 \pm 0.35$ & \nodata & $-6.50 \pm 0.30$ \\
Ca & $-6.96 \pm 0.34$ & $-6.10 \pm 0.13$ & $-5.66 \pm 0.04$ \\
Sc & $-9.76 \pm 0.39$ & $-9.42 \pm 0.06$ & $-8.85 \pm 0.04$ \\
Ti & $-7.86 \pm 0.18$ & $-7.50 \pm 0.07$ & $-7.05 \pm 0.05$ \\
Cr & $-7.59 \pm 0.29$ & $-7.17 \pm 0.08$ & $-6.36 \pm 0.04$ \\
Mn & $-7.65 \pm 0.30$ & $-7.55 \pm 0.08$ & $-6.57 \pm 0.04$ \\
Fe & $-5.44 \pm 0.19$ & $-5.24 \pm 0.03$ & $-4.50 \pm 0.04$ \\
Co & $-8.01 \pm 0.09$ & $-7.79 \pm 0.10$ & $-7.01 \pm 0.07$ \\
Ni & $-7.10 \pm 0.51$ & $-6.68 \pm 0.03$ & $-5.78 \pm 0.04$ \\
Cu & $-8.45 \pm 0.38$ & $-8.73 \pm 0.18$ & $-7.81 \pm 0.04$ \\
Ga & $-8.59 \pm 0.36$ & \nodata & $-8.96 \pm 0.09$ \\
Pd & $-10.00 \pm 0.10$ & \nodata & $-10.43 \pm 0.10$ \\
In & $-11.31 \pm 0.43$ & \nodata & $-11.20 \pm 0.20$ \\
Sn & $-9.88 \pm 0.43$ & \nodata & $-9.96 \pm 0.10$ \\
Hg & $-10.73 \pm 0.39$ & \nodata & $-10.83 \pm 0.08$ \\
Pb & $-10.48 \pm 0.53$ & \nodata & $-10.25 \pm 0.10$ \\
\enddata
\tablecomments{Abundances relative to hydrogen: \abund{X}.  LTE cluster values for C, N, O, Si, and Ca from \citet{Meszaros:2020} and for Sc, Cr, Mn, Co, Ni, and Cu from \citet{Thygesen:2014}.  Non-LTE cluster values for Al from \citeauthor{Thygesen:2014}, for S from \citet{Duffau:2017}, and for Mg, Ti, and Fe from \citet{Kovalev:2019}.}
\end{deluxetable}

\section{Details of the Models}\label{sec_models_discussion}

\subsection{Balmer Lines}\label{sec_halpha_discussion}

Our models cannot reproduce the low-order hydrogen Balmer line profiles.  The model features are broad and shallow, while the observed features have narrow wings and deep cores.  This discrepancy reflects the fact that the low-order lines are formed farther out in the stellar atmosphere, where the low particle density facilitates photoionization, while the high-order lines are formed deeper in the atmosphere, where collision processes dominate and the assumption of LTE is more appropriate.  

\begin{figure}
\plotone{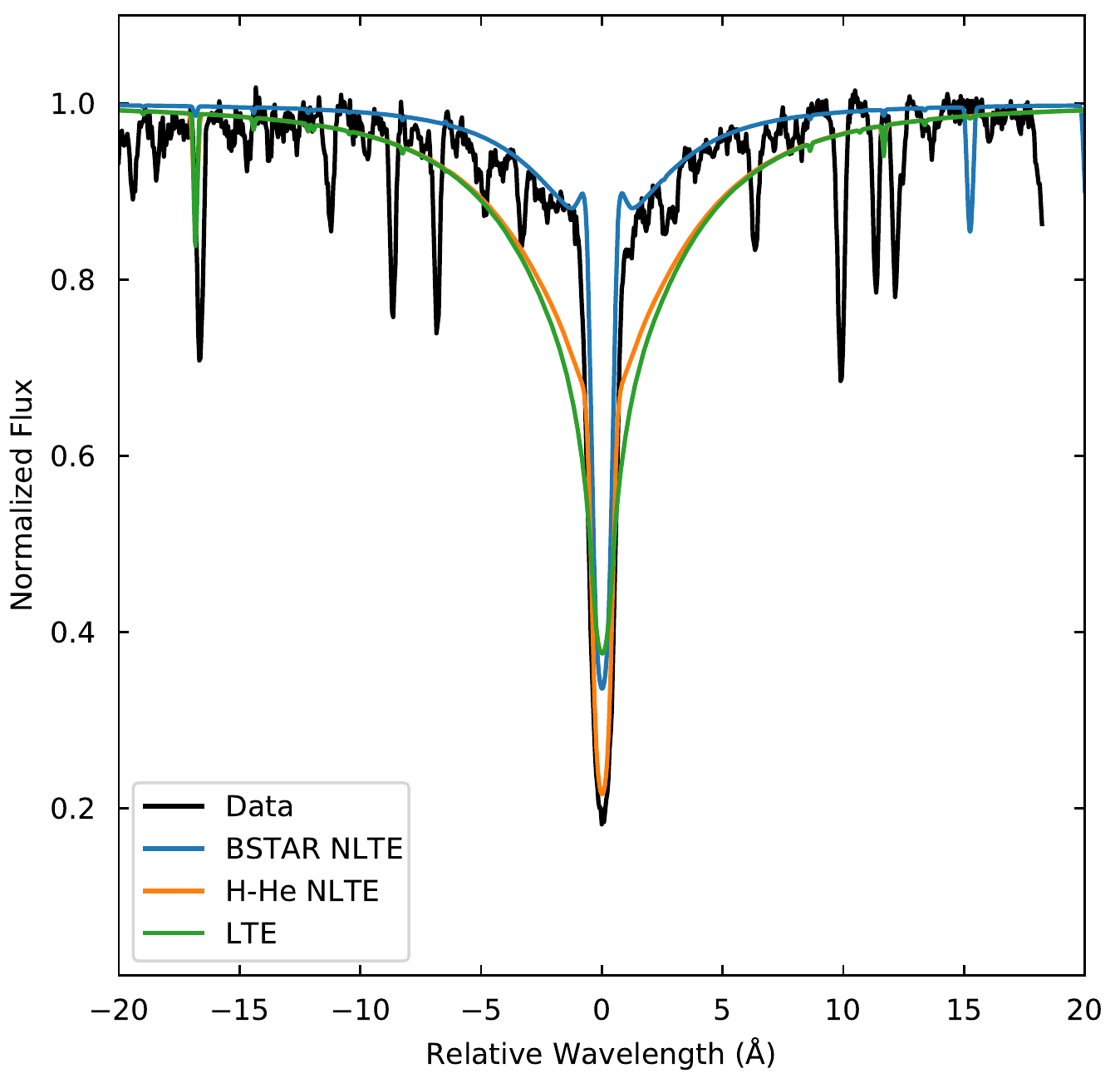}
\caption{The H$\alpha$ line profile.  The black spectrum represents the data, and colored spectra are various models, as described in the text.}
\label{fig_halpha}
\end{figure}

In \figref{fig_halpha}, we plot the H$\alpha$ profile of the Bright Star, together with three synthetic spectra.  The green curve, reproduced from \figref{fig_balmer}, is our best-fit LTE model.  The orange curve represents a simple, non-LTE model computed using the program TLUSTY \citep{Hubeny:Lanz:95}.  The model treats H and He in non-LTE, treats CNO in LTE, and neglects heavier elements.  Metal lines are included in the synthetic spectrum assuming our best-fit abundances.  We see that this model spectrum reproduces the deep, narrow core of the H$\alpha$ profile, but over-predicts the wings of the line.  The blue curve represents a model from the BSTAR grid of \citet{Lanz:Hubeny:2007}, a set of metal line-blanketed, non-LTE, plane-parallel, hydrostatic model atmospheres with parameters appropriate for B-type stars.  Unfortunately, the coolest BSTAR models have effective temperatures \teff\ = 15,000 K, considerably hotter than the Bright Star.  We have selected the model with \teff\ = 15,000 K and \logg\ = 2.25, parameters closest to those of our target.  This model does an excellent job of reproducing the general shape of the line, but the line core is too shallow, because the model is too hot.  These results suggest that a line-blanketed, non-LTE model with the appropriate stellar parameters could reproduce the Bright Star's H$\alpha$ profile.

\subsection{Lyman Lines}\label{sec_lyman_discussion}

The Bright Star's Ly$\alpha$ profile is remarkably broad.  To reproduce it, we employ an ISM absorption model with an \hone\ column density log $N$(\hone) = 21.90 [cm$^{-2}$].  The actual column density toward 47~Tuc is log $N$(\hone) = 20.62 \citep{Heiles:Cleary:1979}.  To confirm this value, we fit the Ly$\alpha$ profile of the star UIT~14 using a model from the OSTAR grid of \citet{Lanz:Hubeny:2003} with \teff\ = 55,000 K, \logg\ = 4.75, and [Fe/H] = $-0.7$.  Using a \citet{Fitzpatrick:1999} extinction curve with \ebv\ = 0.04 and an \hone\ absorption model with log $N$(\hone) = 20.62, we are able to reproduce the observed line profile.  Why then must we assume a higher \hone\ column density to reproduce the Bright Star's Ly$\alpha$ line?

\begin{figure}
\plotone{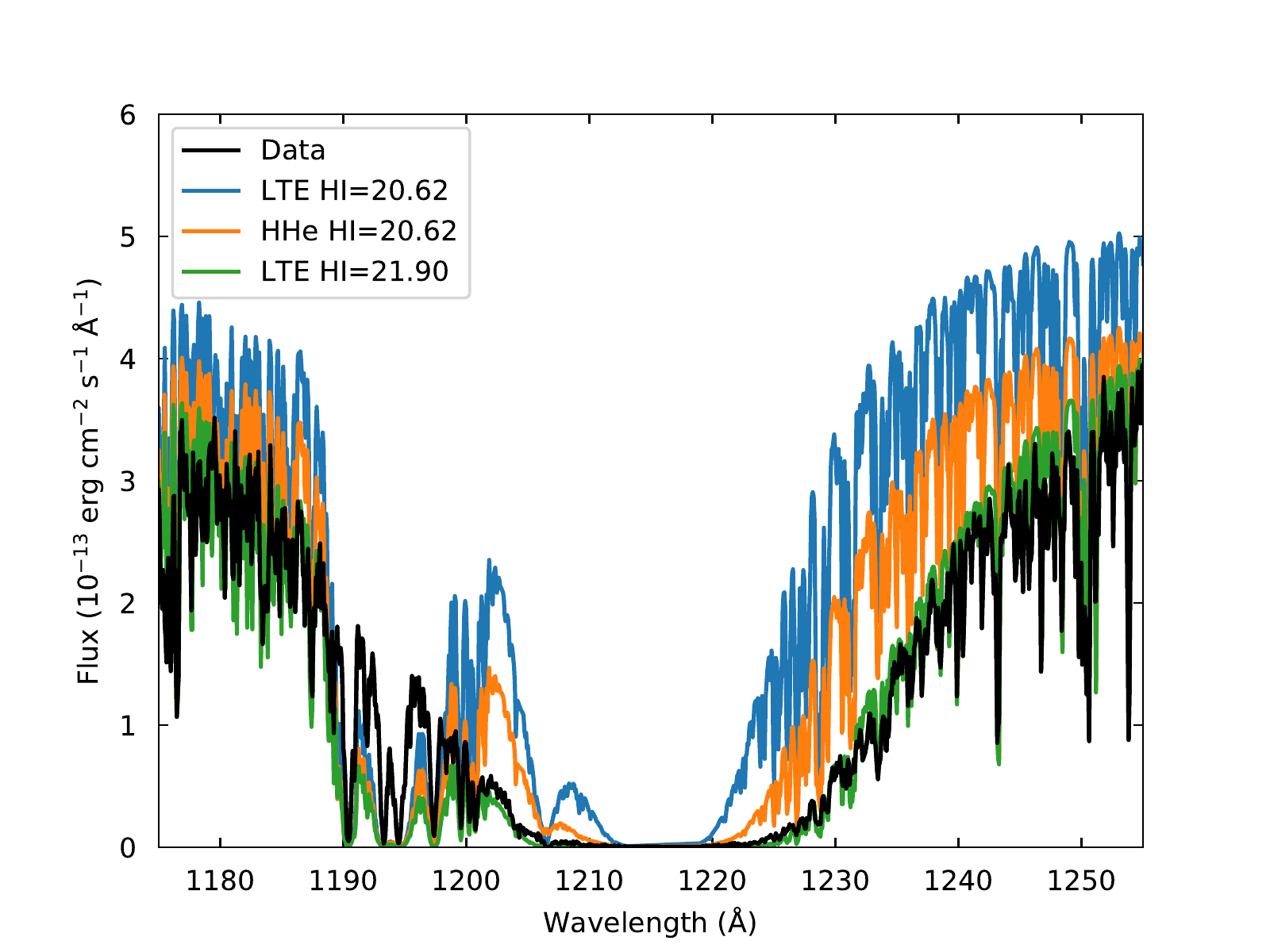}
\caption{The Ly$\alpha$ line profile.  The black spectrum represents the data, and colored spectra are various models, as described in the text.  Both data and models are binned by 7 pixels, the approximate width of a resolution element.  The models are scaled to have the same flux at 1500 \AA.} 
\label{fig_lalpha}
\end{figure}

We can learn something from \figref{fig_lalpha}, where the observed Ly$\alpha$ profile is plotted, along with three model spectra.  The green spectrum is our best-fit model, which employs the high \hone\ column.  It provides a reasonable fit to the data, despite its {\em ad hoc} construction.  The blue spectrum is the same model, but assuming the correct \hone\ column.  Its Ly$\alpha$ feature is far too narrow.  The orange spectrum is derived from the H-He non-LTE model discussed above, again assuming the correct \hone\ column.  Its Ly$\alpha$ profile is broader than that of the LTE model but is still too narrow.  We suspect that additional non-LTE effects broaden the Ly$\alpha$ line.  For example, all of the models over-predict the strength of the \sitwo\ lines at 1190, 1193, 1194, and 1197 \AA\ and under-predict the strength of the \sithree\ line at 1206 \AA, suggesting that these transitions suffer from non-LTE effects.

\begin{figure}
\plotone{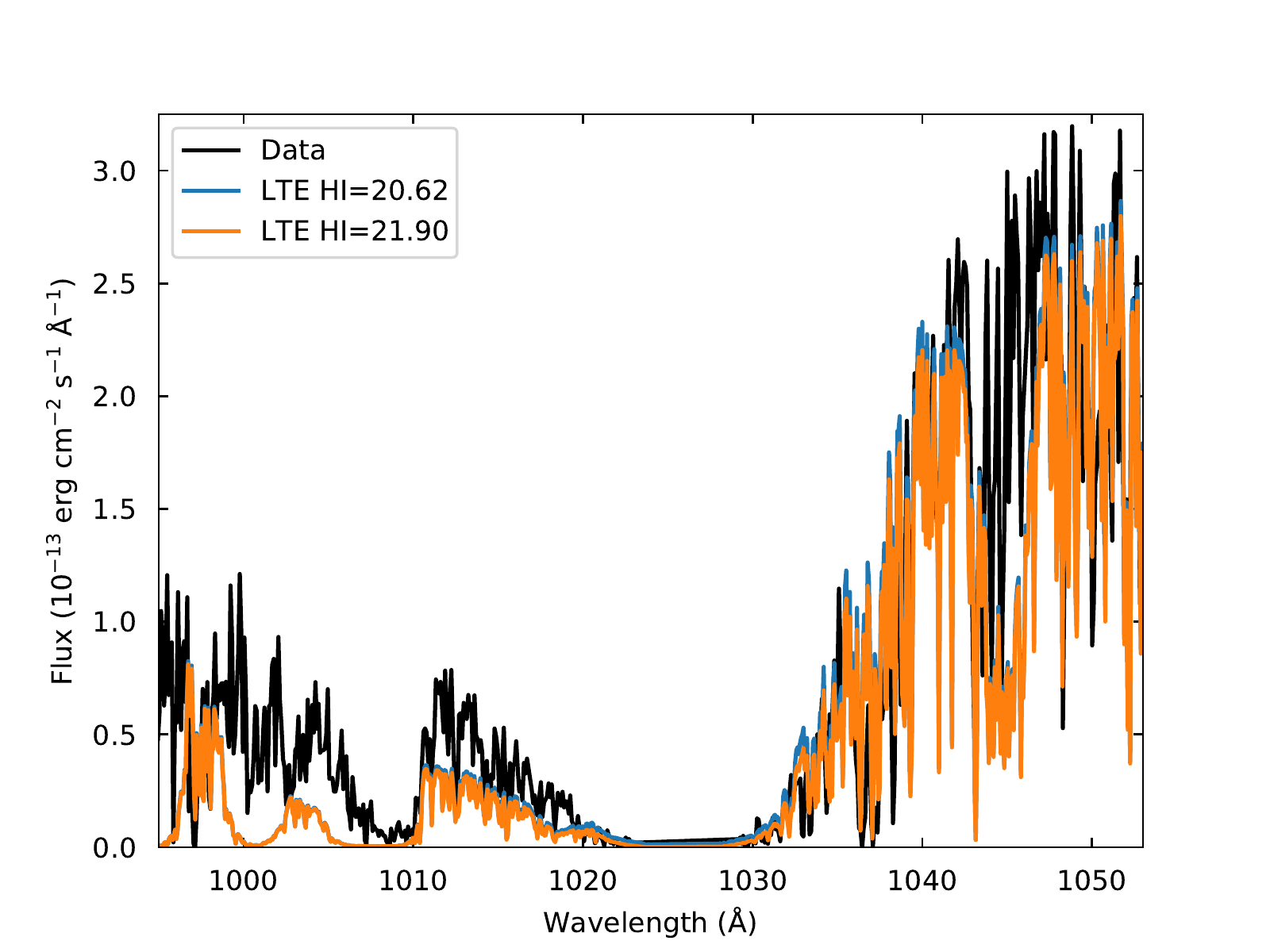}
\caption{Same as \figref{fig_lalpha}, but for Ly$\beta$.} 
\label{fig_lbeta}
\end{figure}

Of the higher-order Lyman lines, only the region around Ly$\beta$ has sufficient flux that we can study its shape.  \figref{fig_lbeta} shows the Ly$\beta$ line profile.  The data are plotted in black, and the blue and orange curves represent our best-fit LTE model with the actual and enhanced \hone\ columns, respectively.  The stellar Ly$\beta$ line is so much broader than the ISM feature that the blue and orange curves are essentially identical.  As shown by \citet{Dixon:Chayer:2013}, the broad troughs in the stellar continuum shortward of Ly$\beta$ are sculpted by resonances in the bound-free absorption cross section of the first excited state of \none.  The broad absorption feature present in the synthetic spectrum near 1045 \AA\ is an absorption resonance in the second excited state of \none.  Both are stronger in our synthetic spectrum than in the data.  Apparently, our LTE model over-predicts the fraction of neutral nitrogen in these excited states.  Even so, our synthetic spectrum reasonably reproduces the Ly$\beta$ line profile.

A more recent measure of the \hone\ column density toward 47~Tuc is provided by the \hone\ 4-PI Survey \citep{HI4PI:2016}, which reports that log $N$(\hone) = 20.75.  This column density yields a Ly$\alpha$ profile slightly wider than that in the spectrum of UIT~14, so we use the lower value of \citet{Heiles:Cleary:1979}.

\subsection{Helium Lines}\label{sec_models_helium}

The \heone\ lines used to derive the star's helium abundance are listed in Table \ref{tab:helium}.  
Strangely, many of these features did not appear in our initial set of synthetic spectra.  The lines were not weaker than expected; they were missing.  SYNSPEC produces an output file that lists all of the lines included in the spectrum, and the \heone\ features did not appear in that file.  After some experimentation, we found that, by adjusting the parameter IDSTD, we were able to restore the missing helium lines.  IDSTD is the index of the standard depth, defined as the depth at which $T \approx$ (2/3)\teff.  For a model with 70 depth points, the nominal value of IDSTD is 50.  Raising IDSTD from 50 to 55 restored the helium features.

\begin{deluxetable}{lcccc}
\tablecaption{Derived Parameters \label{tab:stellar_parms}}
\tablehead{
\colhead{Parameter} & \colhead{Value}
}
\startdata
\teff\ (K) & $10,850  \pm 250$ \\
\logg\ [cm s$^{-2}$] & $2.20 \pm 0.13$ \\
$R_*/R_{\sun}$ & $9.63 \pm 0.13$ \\
$M_*/M_{\sun}$ & $0.54 \pm 0.16$ \\
$\log L_*/L_{\sun}$ & $3.06 \pm 0.04$ \\
\enddata
\end{deluxetable}

\section{Discussion}\label{sec_discussion}

\subsection{Cluster Membership}\label{sec_membership}

Our use of the Bright Star to explore theories of stellar evolution and the chemistry of 47~Tuc is predicated on the assumption that the star is a cluster member.  That membership has been questioned in the literature.  By fitting \citet{Kurucz:79} models to an \iue\/ spectrum of the star, \citet{Thejll:1992} derived a surface gravity (\logg\ = 3.8--4.2, depending on the metallicity) that implies a stellar mass between 71 and 200 \msun\ if the star lies at the distance of the cluster.  He concluded that the star is a Population I foreground star.  

\citet{Dixon:95} derived a lower gravity, \logg\ = 2.0 from the star's H$\beta$ Balmer line and \logg\ = 2.5 from its HUT spectrum, corresponding to a mass between 0.33 and 1.04 \msun.  They also pointed out that, while \iue\/ spectra sample wavelengths longer than than $\sim$ 1150 \AA, HUT spectra extend to the Lyman limit.  In the region between Lyman $\beta$ and Lyman $\alpha$, solar-metallicity models under-predict the observed flux.  The authors concluded that the star does not have Pop I abundances.

We have measured the star's radial velocity using 20 absorption features scattered across the COS spectrum.  We find a mean heliocentric velocity of $-14.3 \pm 2.8$ \kms.  The accuracy of the COS wavelength solution is about 7.5 \kms\ \citep{Plesha:2019}, so the total uncertainty is about 8 \kms.  The cluster has a mean heliocentric velocity of $-18.0$ \kms, with a central velocity dispersion of 11.0 \kms\ \citep{Harris:96, Harris:2010}.  The star's radial velocity is thus consistent with cluster membership.  Finally, inspection of Table \ref{tab:abundance} and \figref{fig:abundance} confirms that the Bright Star's chemical abundances are consistent with those of 47~Tuc.  We conclude that the star is indeed a cluster member.

\subsection{Stellar Mass and Luminosity}\label{sec_mass}

We can derive a star's radius, and from this its mass and luminosity, by comparing its observed and predicted fluxes.  The spectral irradiance of the Bright Star, as reported in {\em Gaia} Early Data Release 3 (EDR3), is $G = 10.952 \pm 0.004$ mag \citep{Gaia_Mission, GaiaEDR3}.  We generate a synthetic spectrum using our best-fit Castelli-Kurucz model, scaled by a \citet{Fitzpatrick:1999} extinction curve with $R_V = 3.1$ and \ebv\ = 0.04 \citep{Harris:96, Harris:2010}.  We compute synthetic stellar magnitudes using the recipe provided in \citet{Riello:2020}.  The ratio between the observed and model fluxes is $\phi = (2.89 \pm 0.01) \times 10^{-20}$.

In the synthetic spectra generated by SYNSPEC, the flux is expressed in terms of the flux moment, $H_\lambda$.  If the star's radius and distance are known, then the scale factor required to convert the model spectrum to the flux at earth is $\phi = 4 \pi (R_* / d)^2$ \citep{Kurucz:79}.  Using {\em Gaia}\/ EDR3 data, \citet{MaizApellaniz:2021} derive a distance to 47~Tuc of $4.53 \pm 0.06$ kpc, in excellent agreement with the eclipsing-binary result of \citet{Thompson:2020}.  Adopting this distance and our scale factor, we derive a stellar radius  $R_*/R_{\sun}$ of $9.63 \pm 0.13$.  Applying our adopted surface gravity (\logg\ = 2.20), we find that the stellar mass $M_*/M_{\sun}$ is $0.54 \pm 0.16$.  Finally, combining the stellar radius with our best-fit effective temperature (\teff\ = 10,850 K), we derive a stellar luminosity $\log L_*/L_{\sun}$ of $3.06 \pm 0.04$.  \citet{Dixon:95} derived a similar value, $\log L_*/L_{\sun}$ of $3.07 \pm 0.20$, from fits to the star's HUT spectrum.

Despite the star's high luminosity, we have identified no wind features in its spectrum.

\begin{figure}
\plotone{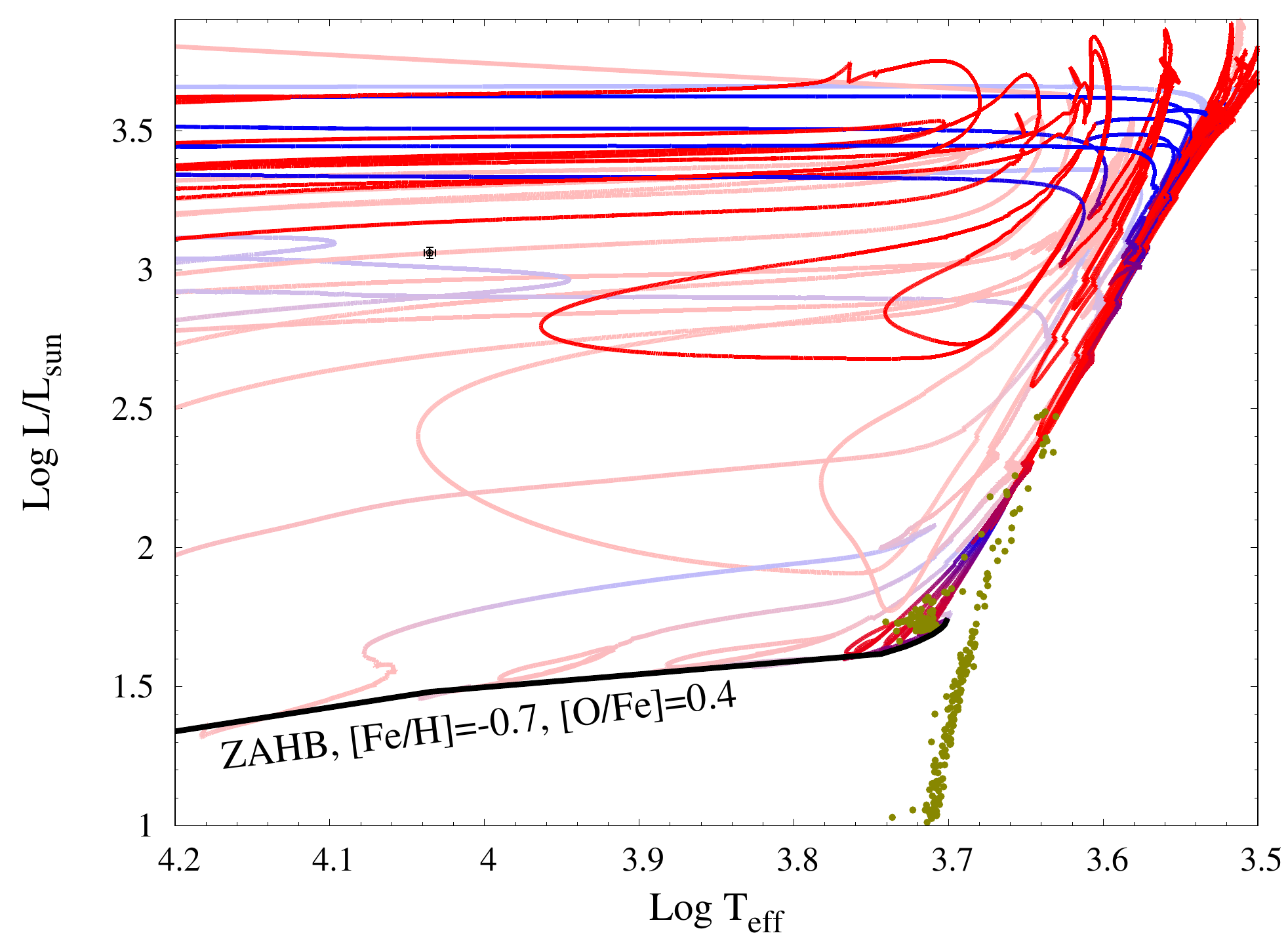}
\caption{Evolutionary tracks for stars similar to those of 47~Tuc during and after the horizontal-branch stage. Darker colors represent more massive objects. The Bright Star is indicated by the black cross. Blue parts of the tracks indicate the regions where energy generation is dominated by H-burning, while red parts indicate the regions where He-burning dominates.}
\label{fig_tracks}
\end{figure}

\subsection{Evolutionary Status}\label{sec_evolution}

To understand the evolutionary history and current status of
the Bright Star in 47~Tuc, we have computed a small grid of stellar-evolution 
models specifically tailored to this cluster. Following the
work of \citet{Denissenkov:2017}, models were computed assuming [Fe/H] =
$-0.7$, [O/Fe] = 0.4, and a helium abundance
$Y=0.27$. Abundance ratios of metals are taken from \citet{Salasnich:2000}.  
This leads to an initial composition of
$(X,Y,Z)=(0.72328,0.27,0.00676)$. The initial mass of the model is
taken as $M_{\rm ZAMS}$ = 0.87 \msun, which corresponds to a total age of
12.76 Gyr for the Bright Star, consistent with the age of
the cluster.  Microphysics and macrophysics prescriptions are taken as
in \citet{Miller_Bertolami:2016}, with the obvious exception of the Rosseland
opacities, which are chosen to reflect the $\alpha$-enhanced
abundances of the cluster. Winds on the RGB are adjusted to populate
the zero-age horizontal branch (ZAHB) as in \citet{Dixon:2017, Dixon:2019}.

The HB and post-HB evolution of the stellar models is illustrated in
\figref{fig_tracks}.  Masses of the tracks
are, from red to blue on the HB ($M_{\rm ZAHB}, M_{\rm
pAGB}$) = (0.87 \msun, 0.554 \msun), (0.80 \msun, 0.547 \msun),
(0.75 \msun, 0.540 \msun), (0.70 \msun, 0.533 \msun), (0.65
\msun, 0.526 \msun), (0.625 \msun, 0.522 \msun), (0.60
\msun, 0.517 \msun), (0.575 \msun, 0.512 \msun), (0.55
\msun, 0.505 \msun), (0.525 \msun, 0.496 \msun), and (0.50
\msun, 0.494 \msun).  The ZAHB of 47~Tuc corresponds to tracks with
masses between 0.65 and 0.80 \msun\
(\teff\ between 5100 K and 5400 K). Darker colors show
the typical evolution of post-red horizontal branch (RHB) sequences
($M \gtrsim$ 0.60 \msun), while lighter tracks show the typical
evolution of post-blue and extreme horizontal branch sequences (BHB
and EHB).

To these evolutionary tracks we have added a sample of cluster stars from the catalog of \citet{Stetson:2019}.  Effective temperatures are estimated from $B-V$ colors, and luminosities from $V$ magnitudes, using the color-color relations and bolometric corrections of \citet{Worthey:2011}.  The cluster distance is taken from \citet{MaizApellaniz:2021} and the reddening from \citet{Harris:96, Harris:2010}.  

47~Tuc hosts a short, red horizontal branch and is essentially devoid
of both BHB and EHB stars. The  Bright Star is thus a decendant of the RHB.  
Colors in \figref{fig_tracks} indicate whether nuclear-energy generation is dominated by
H-burning (blue) or by He-burning (red).
Whether the model departs the AGB as a H-burner or a He-burner
depends on the timing of the last thermal pulse. Models departing 
the AGB during the quiescent H-burning burning phase (between
thermal pulses) will evolve as H-burners, while models departing during or
immediately after a thermal pulse will evolve as He-burners. 

The timing of the last thermal pulse also affects the exact shape of the
loops traced by the star during its post-AGB evolution, as that shape is quite sensitive to
the remaining envelope mass.  Given the uncertainties in the mass-loss
prescriptions and the various processes that may contribute to the
actual wind intensity of the AGB, the phase of the thermal pulse cycle
in which the star departs from the AGB can be considered effectively
stochastic, and a star of a given initial mass can depart at any
phase. As a consequence, the red loops in \figref{fig_tracks} should be considered
representative of the possible trajectories of He-burners within a
given mass range rather than the exact evolution a star with a given
mass.  

It is clear from \figref{fig_tracks} that the Bright Star is less
bright (by about 0.3 dex) than is predicted for post-AGB
H-burners (dark-blue tracks).  In fact, Bright Star's location in the HR
diagram suggests that it is a He-burner. Interestingly, the sequences that evolve
through the observed HB of 47~Tuc ($0.65 M_\odot\lesssim
M_{\rm ZAHB}\lesssim 0.80 M_\odot$) have post-AGB masses ($0.526
M_\odot \lesssim M_{\rm pAGB} \lesssim 0.547 M_\odot$) that are
consistent with the mass of the Bright Star ($0.54 \pm 0.16 M_\odot$).
It is unlikely to be a post-RGB star, as it shows no sign of a binary companion, 
and post-RGB stars with the luminosity of the Bright Star would have significantly lower masses ($\sim 0.4 M_\odot$).

The integrated mass loss between the ZAHB and the post-AGB phase in an
old cluster with $\alpha$-enhanced composition is of special interest, as it
can shed light on the initial-final mass function of these
populations and on the dependence of cold winds on chemical
composition. Reliable determinations of the final stellar masses in globular
clusters (via spectroscopy)
are difficult and consequently rare \citep[\eg,][]{Moehler:2004, Kalirai:2009}, and no previous
determination is available for 47 Tuc.\footnote{From fits to the
cooling sequence of 47~Tuc, \citet{Campos:2016} estimate that young white dwarfs in 47~Tuc have
masses around $0.52 M_\odot$.}  In this context, the mass of the Bright Star is
highly valuable, as it effectively indicates the final mass of single
stars that are now evolving away from the main sequence. 
The main-sequence turn-off mass for 47~Tuc, as inferred from theoretical
isochrones, is $\sim 0.9 M_\odot$,
while the masses of stars on its horizontal branch range between
$\sim 0.65 M_\odot$ and $\sim 0.73 M_\odot$ \citep{Salaris:2016}.
The mass derived for the Bright Star indicates that the stars in 47~Tuc lose
$\Delta M_{\rm AGB} \sim 0.1 M_\odot$--$0.2 M_\odot$ on the AGB, a
value that is only slightly lower than the mass that the same stars lose
on the RGB, $\Delta M_{\rm AGB}\sim 0.17$--$0.21 M_\odot$ \citep{Salaris:2016}.

Low-mass models such as
those presented in this section do not experience third dredge up on
the AGB, and no surface enrichment is expected at this stage. The
N enhancement observed in the Bright Star might be a hint of non-convective mixing
processes, such as rotation-induced mixing or fingering convection, during previous evolutionary stages
\citep{Carbonnel:2010}.

\subsection{Photospheric Abundances}\label{sec_abundance_discussion}

Galactic globular clusters host multiple stellar populations. First-generation (FG) stars display abundances typical of halo field stars, while second-generation (SG) stars, which may have multiple subpopulations, are enriched in He, N, Na, and Al and depleted in C, O and Mg.  Models suggest that the SG stars were formed from gas polluted by material expelled by massive FG stars.  For details, see the review by \citet{Bastian:Lardo:2018}.

\citet{Milone:2012} found that FG stars constitute $\sim$ 30\% of the stars in 47~Tuc.  The majority of the cluster consists of SG stars, which are enhanced in He and N and depleted in C and O.  An additional subpopulation of SG stars, making up about 8\% of the cluster's stars, is visible only on the subgiant branch.  To which generation does the Bright Star belong?  47~Tuc exhibits the Na--O anti-correlation characteristic of all globular clusters.  Several authors use this relationship to define the cluster's subpopulations \citep[\eg,][]{Cordero:2014, Thygesen:2014}.  Since we do not know the Na abundance of the Bright Star, we employ the anti-correlation of the cluster's C and N abundances.

\begin{figure}
\plotone{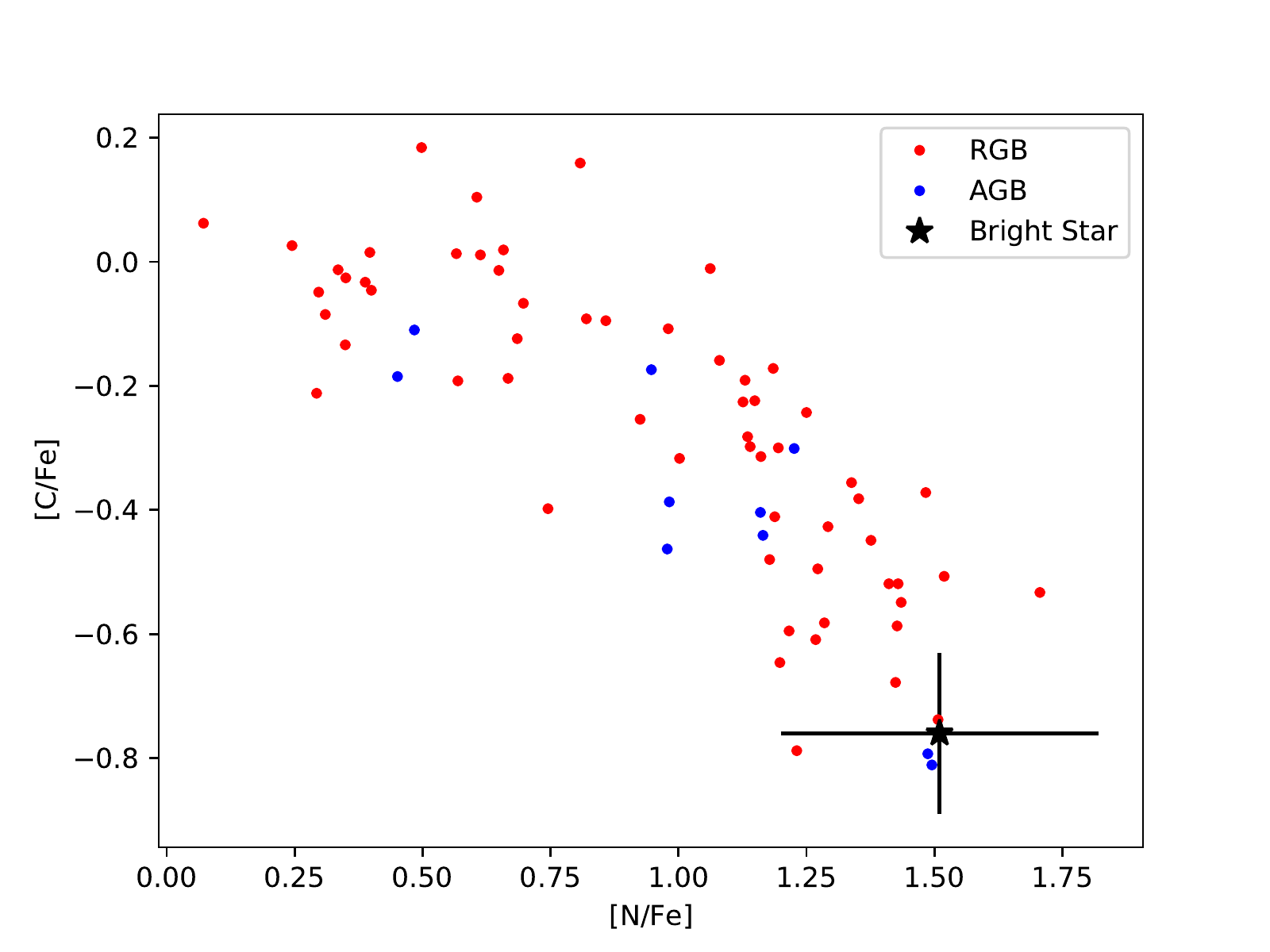}
\caption{Carbon versus nitrogen abundance for stars in 47~Tuc.  Cluster data are from \citealt{Meszaros:2020}.  The Bright Star is indicated by the black symbol.} 
\label{fig_CN}
\end{figure}

In \figref{fig_CN}, we plot the C and N abundances of stars in 47~Tuc, along with those of the Bright Star.  The cluster sample includes both RGB and AGB stars from the full length of the giant branch \citep{Meszaros:2020}.  We see that the Bright Star lies among the most nitrogen-rich, carbon-poor stars in the cluster.  This analysis is complicated by the fact that, within globular clusters, C and N abundances are determined by two astrophysical processes, pollution from FG stars and deep mixing, a non-convective mixing process that results in a steady depletion of carbon and an enhancement of nitrogen as stars evolve up the RGB \citep{Gratton:2004}.  If deep mixing were the dominant effect, we would expect all of the cluster's AGB stars---which presumably evolved to the tip of the RGB---to be enhanced in N; instead, they show the same range of N abundance as the RGB stars.  We conclude that, while the Bright Star's C/N ratio may have been reduced as it climbed the RGB, it began life as a SG star.

Returning to \figref{fig:abundance}, we see that abundances of the intermediate-mass elements (Mg through Ga) generally scale with Fe.  Where cluster values are available, all but copper are slightly higher than those measured for the Bright Star.  Though the difference is small for any single element, the uniformity of this offset is striking.  It suggests that systematic effects, whether the use of different model atmospheres, absorption features, or temperature regimes, may be at play.

The heaviest elements, Pd through Pb, have roughly solar abundances.  These elements are generally associated with s-process nucleosynthesis on the AGB; however, the star's low carbon abundance ($N_{\rm C} \ll N_{\rm O}$) indicates that it did not undergo third dredge-up while on the AGB and implies that the heaviest elements were made by a previous generation of stars.  If so, then we would expect this pattern to be present throughout the cluster, not just in this star.

\begin{figure}
\plotone{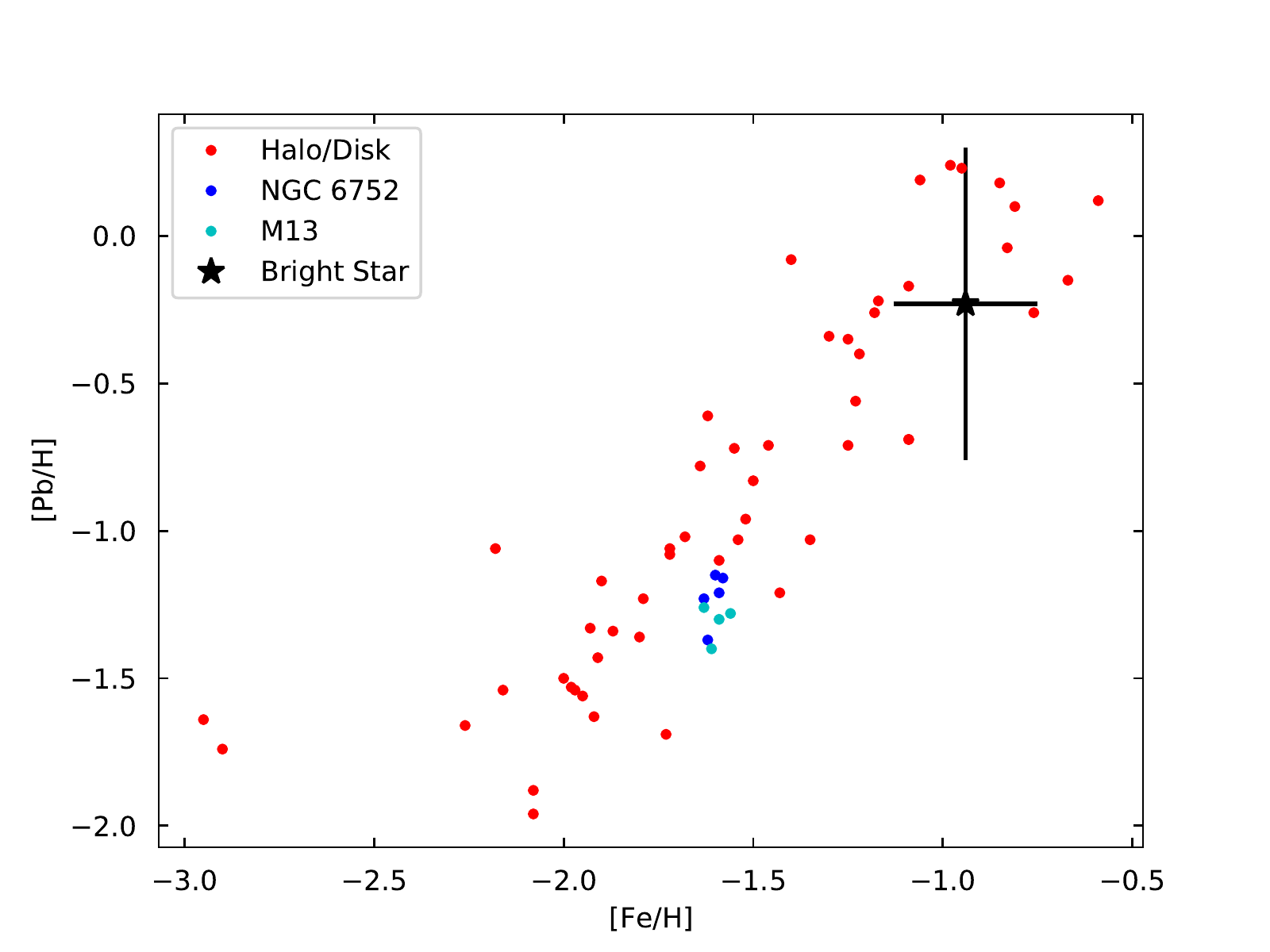}
\caption{Lead abundance versus metallicity for metal-poor dwarf and giant stars in the Galactic halo and disk and giant stars in NGC~6752 and M13.  The Bright Star is indicated by the black symbol.} 
\label{fig_lead}
\end{figure}

\citet{Roederer:2010} derived the Pb abundance of a sample of metal-poor dwarf and giant stars in the Galactic halo and disk by fitting LTE models to their \pbone\ $\lambda 4057$ lines.  The resulting abundances are subject to LTE effects, because Pb is mostly ionized in stars with \teff\ $> 4000$ K.  \citet{Mashonkina:2012} derived non-LTE corrections for these stars.  In \figref{fig_lead}, we plot the corrected value of [Pb/H] as a function of [Fe/H] from \citeauthor{Mashonkina:2012}  \citet{Yong:2006} measured the Pb abundance in five giant stars in the globular cluster NGC~6752 and four in M13, again fitting LTE models to the \pbone\ $\lambda 4057$ line.  We apply the \citeauthor{Mashonkina:2012} corrections to their results and plot them as well.  Finally, we plot the Bright Star.  Because its Pb abundance was derived from a \pbtwo\ line, non-LTE effects should be small.  We see that the stars in NGC~6752, M13, and 47~Tuc follow the same trend as the halo and disk stars.

\section{Conclusions}\label{sec_conclusions}

We have analyzed FUV and optical spectra of the Bright Star in 47 Tuc.  Its high-order Balmer lines, H06 and above, are well-fit by LTE stellar-atmosphere models, but the low-order lines are not.  These features are predicted to be broad and shallow, while the observed features have narrow wings and deep cores.  Comparisons with metal line-blanketed non-LTE models at higher effective temperatures suggest that similar models at lower temperatures could reproduce the observed features.

The horizontal branch of 47~Tuc is short and red.  Stellar-evolution models that reproduce the cluster's ZAHB have post-AGB masses consistent with that of the Bright Star.  The star's relatively low luminosity indicates that it is now a He-burner, having left the AGB during or immediately after a thermal pulse.  Its derived mass suggests that single stars in 47~Tuc lose $\Delta M_{\rm AGB} \sim 0.1 M_\odot$--$0.2 M_\odot$ on the AGB, a value that is only slightly lower than the mass the same stars lose on the RGB.

The star's low C/N ratio suggests that it belongs to the second generation of cluster stars, while its low C/O ratio indicates that it did not undergo third dredge-up on the AGB.  Abundances of intermediate-mass elements are consistent with cluster values, while heavy-element abundances are essentially solar.  We conclude that the star experienced no significant change in its photospheric abundances as it climbed the AGB.  If so, then its heavy-element abundances are typical of the cluster values.  Thus, the Bright Star provides a unique look at the abundances of heavy elements in 47~Tuc.

\acknowledgments

The authors would like to think Christian Johnson for helpful comments on the manuscript.
P.C.\ is supported by the Canadian Space Agency under a contract with NRC Herzberg Astronomy and Astrophysics.
This work has made use of
NASA's Astrophysics Data System (ADS); 
the SIMBAD database, operated at CDS, Strasbourg, France; 
the Mikulski Archive for Space Telescopes (MAST), hosted at the Space Telescope Science Institute, which is operated by the Association of Universities for Research in Astronomy, Inc., under NASA contract NAS5-26555; 
data, software and/or web tools obtained from the High Energy Astrophysics Science Archive Research Center (HEASARC), a service of the Astrophysics Science Division at NASA/GSFC and of the Smithsonian Astrophysical Observatory's High Energy Astrophysics Division; and
data from the European Space Agency (ESA) mission
{\it Gaia} (\url{https://www.cosmos.esa.int/gaia}), processed by the {\it Gaia}
Data Processing and Analysis Consortium (DPAC,
\url{https://www.cosmos.esa.int/web/gaia/dpac/consortium}). 
Funding for the DPAC has been provided by national institutions, 
in particular the institutions participating in the {\it Gaia} Multilateral Agreement.
Publication of this work is supported by the STScI Director's Discretionary Research Fund.

%

\vspace{5mm}
\facilities{FUSE, HST (COS), Magellan:Clay (MIKE spectrograph)}

\clearpage

\appendix

\restartappendixnumbering

\section{Selected Features in the Bright Star's Spectrum}

\startlongtable
\begin{deluxetable*}{lcDrcc}
\tablecaption{Selected FUV Absorption Features \label{tab:lines_fuv}}
\tablehead{
\colhead{Ion} & \colhead{$\lambda_{\rm lab}$} & \multicolumn2c{$\log gf$} & \colhead{$E_l$} & \colhead{Abundance} & \colhead{Grating} \\
\colhead{} & \colhead{(\AA)} & \multicolumn2c{} & \colhead{(cm$^{-1}$)}
}
\decimals
\startdata
\cone  &  1328.834  &  -1.20  &   0.000   & $ -5.90 \pm 0.19 $ & G130M \\
       &  1329.085  &  -1.20  &  16.400   & \nodata & \\
       &  1329.100  &  -1.11  &  16.400   & \nodata & \\
       &  1329.123  &  -1.32  &  16.400   & \nodata & \\
       &  1329.577  &  -0.62  &  43.400   & $ -6.22 \pm 0.19 $ & \\
       &  1329.600  &  -1.10  &  43.400   & \nodata & \\
       &  1656.267  &  -0.75  &  16.400  & $ -6.35 \pm 0.20 $ & G160M \\
       &  1657.907  &  -0.85  &  16.400  & $ -6.13 \pm 0.18 $ & \\
       &  1658.121  &  -0.75  &  43.400  & $ -6.21 \pm 0.18 $ & \\
\ctwo  &  1323.862  &  -1.30  &  74930.100   & $ -5.26 \pm 0.13 $ & G130M \\
       &  1323.906  &  -0.34  &  74932.620   & \nodata & \\
       &  1323.951  &  -0.15  &  74930.100   & \nodata & \\
       &  1323.995  &  -1.30  &  74932.620   & \nodata & \\
\none  &  1165.594  & -2.56 &  19224.465  & $ -3.93 \pm 0.21 $ & G130M \\
       &  1165.717  & -3.99 &  19224.465  & \nodata &  \\
       &  1165.836  & -3.61 &  19233.178  & \nodata & \\
       &  1169--1173\tablenotemark{a} & \nodata &  \nodata\phn\phn  & $ -3.62 \pm 0.14 $ \\
       &  1176.510  & -1.15 &  19224.465  & $ -3.41 \pm 0.21 $ & \\
       &  1177.695  & -1.37 &  19233.177  & $ -4.57 \pm 0.34 $ &  \\
       &  1243.171  & -1.51 &  19224.465  & $ -3.48 \pm 0.22 $ & \\
       &  1243.179  & -0.35 &  19224.465  & \nodata & \\
       &  1243.306  & -0.54 &  19233.177  & \nodata & \\
       &  1243.313  & -1.49 &  19233.177  & \nodata & \\
       &  1318.823  & -2.55 &  28838.919  & $ -3.81 \pm 0.25 $ & \\
       &  1318.998  & -1.53 &  28838.919  & \nodata & \\
       &  1319.005  & -1.84 &  28839.307  & \nodata & \\
       &  1319.669  & -1.68 &  28838.919  & $ -4.16 \pm 0.34 $ & \\
       &  1319.676  & -1.18 &  28839.307  & \nodata & \\
       &  1327.917  & -2.42 &  28838.919  & $ -3.27 \pm 0.36$ & \\
       &  1327.924  & -2.72 &  28839.307  & \nodata  \\
\ntwo  &  1300.035  & -5.25 &  15316.200  & $ -3.43 \pm 0.37 $ & G130M \\
\oone  &  1152.151  & -0.28 &  15867.862    & $-3.60 \pm 0.13$ & \fuse\ \\
       &  \multicolumn4c{}    & $-3.61 \pm 0.16$ & G130M \\
\mgtwo  &  1307.875  & -2.49 &  35669.309  & $ -4.58 \pm 0.23 $ & G130M \\
        &  1365.544  & -2.33 &  35669.309  & $ -5.25 \pm 0.18 $ &  \\
        &  1367.708  & -2.24 &  35669.309  & $ -5.27 \pm 0.20 $ &  \\
        &  1369.423  & -1.94 &  35760.880  & $ -4.96 \pm 0.29 $ &  \\
        &  1398.776  & -4.83 &      0.000  & $ -4.73 \pm 0.23 $ &  \\
        &  1476.000  & -1.84 &  35669.309  & $ -4.91 \pm 0.32 $ & G160M \\
        &  1482.890  & -1.62 &  35760.880  & $ -5.02 \pm 0.35 $ &  \\
        &  1734.852  & -1.10 &  35669.309  & $ -5.16 \pm 0.48 $ &  \\
        &  1737.613  & -1.81 &  35760.880  & $ -5.46 \pm 0.37 $ &  \\
        &  1737.628  & -0.85 &  35760.880  & \nodata &  \\
        &  1753.474  & -1.15 &  35760.880  & $ -4.67 \pm 0.40 $ &  \\
\altwo  &  1670.787  &   0.32  &  0.000  & $ -6.20 \pm 0.09 $ & G160M \\
\sitwo  &  1264.738  &  +0.55  &    287.240  & $ -5.09 \pm 0.05 $ & G130M \\
        &  1265.002  &  -0.41  &    287.240  & \nodata & \\
        &  1309.276  &  -0.12  &    287.240  & $ -5.15 \pm 0.12 $ & \\
        &  1309.453  &  +0.55  &  55309.353  & \nodata & \\
        &  1346.884  &  -0.55  &  42932.622  & $ -4.23 \pm 0.28 $ \\
        &  1348.543  &  -0.57  &  42824.289  & $ -4.90 \pm 0.34 $ \\
        &  1350.072  &  -0.17  &  43107.910  & $ -4.87 \pm 0.30 $ \\
        &  1350.516  &  -1.07  &  42932.622  & $ -4.91 \pm 0.27 $ \\
        &  1350.656  &  -1.27  &  42824.289  & \nodata &  \\
        &  1353.721  &  -0.53  &  43107.910  & $ -4.94 \pm 0.27 $ &  \\
        &  1533.431  &  -0.29  &    287.240  & $ -5.12 \pm 0.08 $ & G160M \\
        &  1710.836  &  -0.50  &  55309.353  & $ -5.41 \pm 0.34 $ & \\
        &  1711.304  &  -0.20  &  55325.181  & $ -5.45 \pm 0.34 $ & \\
\sithree &  1296.726  &  -0.13  &  52724.691  & $ -4.85 \pm 0.12 $ & G130M \\
        &  1301.149  &  -0.13  &  52853.283  & $ -4.90 \pm 0.11 $ \\
        &  1417.237  &  -0.11  &  82884.404  & $ -4.93 \pm 0.22 $ \\
\ptwo  &  1149.958  &  -0.51  &  164.900  & $ -7.34 \pm 0.45 $ & G130M \\
       &  1155.014  &  -0.74  &  164.900  & $ -7.30 \pm 0.64 $ & \\
       &  1310.703  &  -1.38  &  469.120  & $ -7.10 \pm 0.36 $ & \\
       &  1542.304  &  -1.45  &  469.120  & $ -7.23 \pm 0.36 $ & G160M \\
       &  1543.133  &  -2.24  &  469.120  & $ -7.09 \pm 0.32 $ &  \\
\sone  &  1425.188  &  -0.87  &     0.000  & $ -6.22 \pm 0.27 $ & G130M \\
       &  1425.219  &  -2.05  &     0.000  & \nodata  &       \\
       &   \multicolumn4c{}      &  $ -6.26 \pm 0.28 $  & G160M \\
       &  1436.967  &  -0.72  &   573.640  & $ -6.00 \pm 0.28 $ &  \\
       &  1448.229  &  -0.19  &  9238.609  & $ -5.88 \pm 0.25 $ &  \\
       &  1473.994  &  -0.43  &     0.000  & $ -6.18 \pm 0.26 $ &  \\
\cltwo  &  1079.080  &  -1.62  &  696.000  & $ -7.13 \pm 0.35 $ & \fuse \\
\catwo  &  1288.210  &  -1.29  &  13710.880  & $ -6.37 \pm 0.32 $ & G130M \\
        &  1553.176  &  -0.35  &  13650.190  & $ -6.95 \pm 0.28 $ & G160M \\
        &  1554.642  &  -0.20  &  13710.880  & $ -7.42 \pm 0.28 $ &  \\
        &  1554.642  &  -1.50  &  13710.880  &  &  \\
\scthree  &  1610.194  &  -0.47  &  0.000  & $ -9.76 \pm 0.39 $ & G160M \\
\tithree  &  1293.225  &  -0.75  &  420.400  & $ -7.99 \pm 0.42 $ & G130M \\
          &  1339.703  &  -1.67  &  8473.500  & $ -7.51 \pm 0.31 $ &  \\
          &  1421.641  &  -0.92  &  10603.600  & $ -7.93 \pm 0.25 $ &  \\
          &  1421.755  &  -1.01  &  10603.600  & $ -7.75 \pm 0.27 $ &  \\
          &  1491.975  &  -1.02  &  10721.200  & $ -7.96 \pm 0.30 $ & G160M \\
          &  1498.695  &  -0.33  &  8473.500  & $ -7.95 \pm 0.40 $ &  \\
          &  1499.177  &  -0.94  &  10721.200  & $ -8.06 \pm 0.20 $ &  \\
          &  1504.623  &  -1.39  &  10538.400  & $ -7.94 \pm 0.31 $ &  \\
          &  1506.101  &  -1.63  &  10603.600  & $ -7.62 \pm 0.36 $ &  \\
\crtwo &  1427.382  &  -0.59  &  12303.860  & $ -7.80 \pm 0.28 $ & G130M \\
       &  \multicolumn4c{}                  & $ -7.75 \pm 0.30 $ & G160M \\
       &  1431.859  &  -0.05  &  12303.860  & $ -7.78 \pm 0.36 $ & G130M \\
       &  \multicolumn4c{}                  & $ -7.86 \pm 0.27 $ & G160M \\
       &  1435.202  &  -0.27  &  12147.821  & $ -7.82 \pm 0.16 $ & G160M \\
       &  1435.229  &  -0.85  &  12032.580  & \nodata \\
       &  1435.251  &  -1.31  &  12303.860  & \nodata \\
\crthree &  1426.985  &  -1.27  &  49768.646  & $ -7.36 \pm 0.50 $ & G130M \\
         &  \multicolumn4c{}                  & $ -7.06 \pm 0.29 $ & G160M \\
\mntwo &  1163.326  &  -1.30  &  0.000  & $ -7.23 \pm 0.39 $ & G130M \\ 
       &  1464.570  &  -0.39  &  14325.860  & $ -8.06 \pm 0.35 $ & G160M \\
       &  1732.700  &  -0.72  &  14593.819  & $ -7.41 \pm 0.30 $ \\
       &  1740.144  &  -0.81  &  14781.190  & $ -7.87 \pm 0.37 $ \\
       &  1741.977  &  -0.86  &  14901.180  & $ -7.68 \pm 0.43 $ \\
\fetwo & 1610--1619\tablenotemark{a} & \nodata &  \nodata\phn\phn  & $ -5.25 \pm 0.29 $ & G160M \\
       & 1620--1629\tablenotemark{a} & \nodata &  \nodata\phn\phn  & $ -5.25 \pm 0.31 $ \\
       & 1630--1640\tablenotemark{a} & \nodata &  \nodata\phn\phn  & $ -5.36 \pm 0.50 $ \\
       & 1640--1650\tablenotemark{a} & \nodata &  \nodata\phn\phn  & $ -5.25 \pm 0.34 $ \\
       & 1650--1660\tablenotemark{a} & \nodata &  \nodata\phn\phn  & $ -5.36 \pm 0.34 $ \\
       & 1660--1669\tablenotemark{a} & \nodata &  \nodata\phn\phn  & $ -5.43 \pm 0.49 $ \\
       & 1672--1680\tablenotemark{a} & \nodata &  \nodata\phn\phn  & $ -5.46 \pm 0.38 $ \\
       & 1680--1690\tablenotemark{a} & \nodata &  \nodata\phn\phn  & $ -5.34 \pm 0.68 $ \\
       & 1690--1700\tablenotemark{a} & \nodata &  \nodata\phn\phn  & $ -5.85 \pm 0.41 $ \\
       & 1701--1709\tablenotemark{a} & \nodata &  \nodata\phn\phn  & $ -5.71 \pm 0.26 $ \\
       & 1710--1720\tablenotemark{a} & \nodata &  \nodata\phn\phn  & $ -5.75 \pm 0.33 $ \\
       & 1720--1730\tablenotemark{a} & \nodata &  \nodata\phn\phn  & $ -5.40 \pm 0.39 $ \\
       & 1730--1740\tablenotemark{a} & \nodata &  \nodata\phn\phn  & $ -5.35 \pm 0.44 $ \\
       & 1755--1760\tablenotemark{a} & \nodata &  \nodata\phn\phn  & $ -5.23 \pm 0.37 $ \\
\cotwo &  1448.011  &  -0.39  &  0.000  & $ -7.97 \pm 0.38 $ & G160M \\
       &  1455.885  &  -0.39  &  950.510  & $ -7.88 \pm 0.40 $ \\
       &  1466.203  &  0.10  &  0.000  & $ -8.16 \pm 0.51 $ \\
       &  1472.902  &  0.01  &  950.510  & $ -8.05 \pm 0.41 $ \\
       &  1476.666  &  -0.51  &  1597.320  & $ -7.97 \pm 0.35 $ \\
\nitwo &  1500.434  &  -0.72  &  1506.940  & $ -7.04 \pm 0.54 $ & G160M \\
       &  1501.959  &  -2.34  &  0.000  & $ -6.52 \pm 0.34 $ \\
       &  1502.148  &  -1.10  &  0.000  & $ -6.60 \pm 0.41 $ \\
       &  1536.741  &  -1.35  &  1506.940  & $ -7.30 \pm 0.38 $ \\
       &  1536.939  &  -1.64  &  1506.940  & $ -6.56 \pm 0.46 $ \\
       &  1703.405  &  -1.13  &  0.000  & $ -7.35 \pm 0.41 $ \\
       &  1748.282  &  -0.25  &  1506.940  & $ -8.16 \pm 0.44 $ \\
       &  1754.809  &  -0.92  &  1506.940  & $ -7.58 \pm 0.47 $ \\
\cutwo &  1358.773  &  -0.27  &  0.000  & $ -8.98 \pm 0.50 $ & G130M \\
       &  1367.951  &  -1.48  &  0.000  & $ -8.29 \pm 0.40 $ &  \\
       &  1472.395  &  -2.08  &  0.000  & $ -8.08 \pm 0.38 $ & G160M \\
\gatwo   &  1414.402  &   0.35  &      0.000  & $ -10.53 \pm 0.31 $ & G130M \\
         &  1473.690  &  -0.24  &  47367.548  & $  -8.49 \pm 0.40 $ & G160M \\
         &  1483.453  &  -0.37  &  47814.115  & $  -8.54 \pm 0.47 $ &  \\
         &  1483.903  &   0.33  &  48749.852  & $  -9.19 \pm 0.57 $ &  \\
         &  1495.185  &  -0.25  &  47814.115  & $  -8.85 \pm 0.64 $ &  \\
         &  1504.331  &  -0.16  &  48749.852  & $  -8.57 \pm 0.41 $ &  \\
         &  1504.926  &  -0.10  &  47367.548  & $  -8.19 \pm 0.31 $ &  \\
         &  1515.106  &  -0.22  &  47814.115  & $  -8.03 \pm 0.32 $ &  \\
         &  1535.309  &   0.53  &  48749.852  & $  -8.90 \pm 0.33 $ &  \\
\gathree &  1495.045  &   0.03  &      0.000  & $  -8.79 \pm 0.40 $ &  \\
         &  1534.462  &  -0.28  &      0.000  & $  -9.15 \pm 0.55 $ &  \\
\pdtwo  &  1345.556  &  0.00  &  0.000  & $ -9.89 \pm 0.34 $ & G130M \\
        &  1363.745  &  0.00  &  0.000  & $ -10.15 \pm 0.32 $ \\
        &  1397.448  &  0.00  &  3539.000  & $ -9.93 \pm 0.35 $ \\
        &  1403.613  &  0.00  &  0.000  & $ -10.04 \pm 0.32 $ \\
\intwo  &  1586.450  &  0.39  &  0.000  & $ -11.31 \pm 0.43 $ & G160M \\
\sntwo &  1475.011  &  0.47  &  4251.494  & $ -9.88 \pm 0.43 $ & G160M \\
\hgtwo  &  1649.947  &  0.22  &  0.000  & $ -10.73 \pm 0.39 $ & G160M \\
\pbtwo  &  1433.905  &  0.24  &  0.000  & $ -10.48 \pm 0.53 $ & G160M \\
\enddata
\tablecomments{Abundances relative to hydrogen: \abund{X}.}
\tablenotetext{a}{Spectral region with multiple features.  See discussion in text.}
\end{deluxetable*}

\startlongtable
\begin{deluxetable*}{llDrrc}
\tablecaption{Selected Optical Absorption Features \label{tab:lines_mike}}
\tablehead{
\colhead{Ion} & \colhead{$\lambda_{\rm lab}$} & \multicolumn2c{$\log gf$} & \colhead{$E_l$} & \colhead{EW (m\AA)} & \colhead{Abundance} \\
\colhead{} & \colhead{(\AA)} & \multicolumn2c{} & \colhead{(cm$^{-1}$)} & \colhead{}
}
\decimals
\startdata
\none  &  7468.312  &  -0.27  &  83364.615 & $ 43.5 \pm 1.3 $ & $ -3.47 \pm 0.02 $ \\
       &  8216.336  &  -0.01  &  83364.615  & $ 63.9 \pm 1.7 $ & $ -3.30 \pm 0.02 $ \\
       &  8223.128  &  -0.39  &  83317.826  & $ 38.7 \pm 1.7 $ & $ -3.42 \pm 0.02 $ \\
       &  8680.282  &   0.24  &  83364.615  & $ 74.2 \pm 2.5 $ & $ -3.71 \pm 0.02 $ \\
       &  8683.403  &  -0.04  &  83317.826  & $ 67.1 \pm 2.3 $ & $ -3.59 \pm 0.02 $ \\
       &  8686.149  &  -0.45  &  83284.068  & $ 51.9 \pm 2.4 $ & $ -3.43 \pm 0.02 $ \\
       &  8703.247  &  -0.41  &  83284.068  & $ 54.4 \pm 2.2 $ & $ -3.46 \pm 0.02 $ \\
       &  8711.703  &  -0.38  &  83317.826  & $ 52.8 \pm 2.0 $ & $ -3.51 \pm 0.02 $ \\
       &  8718.837  &  -0.43  &  83364.615  & $ 51.6 \pm 2.0 $ & $ -3.47 \pm 0.02 $ \\
\ntwo  &  3994.997  &  0.28  &  149187.793  & $ 18.0 \pm 2.0 $ & $ -3.43 \pm 0.05 $ \\
       &  4630.539  &  0.13  &  149076.512  & $  9.7 \pm 0.9 $ & $ -3.45 \pm 0.05 $ \\
\oone  &  6156.778  &  -0.73  &  86627.782    & $8.8 \pm 1.4$ & $-3.97 \pm 0.04$ \\
       &  6158.187  &  -0.44  &  86631.453    & $12.7 \pm 1.3$ & $-3.97 \pm 0.03$ \\
       &  7771.944  &   0.32  &  73768.202  & $79.1 \pm 1.5$ & $-3.87 \pm 0.02$ \\
       &  7774.166  &   0.17  &  73768.202  & $75.6 \pm 1.5$ & $-3.78 \pm 0.02$ \\
       &  7775.388  &  -0.05  &  73768.202  & $64.2 \pm 1.5$ & $-3.76 \pm 0.02$ \\
       &  7950.803  &   0.34  &  101147.522   & $10.1 \pm 1.5$ & $-3.60 \pm 0.04$ \\
       &  7952.159  &   0.17  &  101155.417   & $8.6 \pm 1.6$ & $-3.53 \pm 0.05$ \\
       &  7981.942  &  -0.47  &  88630.585  & $13.1 \pm 2.1$ & $-3.93 \pm 0.04$ \\
       &  7982.398  &  -0.34  &  88631.307  & \nodata\phn\phn & \nodata \\
       &  8446.247  &  -0.52  &  76794.977    & $148.1 \pm 2.4$ & $-3.72 \pm 0.01$ \\
       &  8446.359  &   0.17  &  76794.977    & \nodata\phn\phn & \nodata \\
       &  8446.758  &  -0.05  &  76794.977    & \nodata\phn\phn & \nodata \\
\neone  &  6096.163  &  -0.27  &  134459.279  & $  7.7 \pm 1.6 $ & $ -4.04 \pm 0.10 $ \\
        &  6334.428  &  -0.31  &  134041.849  & $  9.8 \pm 1.4 $ & $ -3.80 \pm 0.06 $ \\
        &  6382.991  &  -0.26  &  134459.279  & $  8.4 \pm 1.3 $ & $ -3.92 \pm 0.06 $ \\
        &  6402.246  &   0.36  &  134041.849  & $ 21.1 \pm 1.4 $ & $ -3.91 \pm 0.04 $ \\
        &  6506.528  &   0.03  &  134459.279  & $ 13.6 \pm 1.3 $ & $ -3.92 \pm 0.04 $ \\
\mgone  &  3832.299  &  -0.36  &  21870.465  & $ 12.8 \pm 1.6 $ & $ -4.89 \pm 0.05 $ \\
        &  3832.304  &   0.12  &  21870.465  & \nodata\phn\phn & \nodata \\
        &  3838.290  &  -1.53  &  21911.178  & $ 17.9 \pm 1.5 $ & $ -4.89 \pm 0.04 $ \\
        &  3838.292  &   0.39  &  21911.178  & \nodata\phn\phn & \nodata \\
        &  3838.295  &  -0.35  &  21911.178  & \nodata\phn\phn & \nodata \\
\mgtwo  &  3848.211  &  -1.59  &  71490.190  & $ 12.1 \pm 2.3 $ & $ -5.04 \pm 0.06 $ \\
        &  3848.341  &  -2.54  &  71491.064  & \nodata\phn\phn & \nodata \\
        &  3850.386  &  -1.84  &  71491.064  & $ 7.6 \pm 2.6 $ & $ -4.98 \pm 0.11 $ \\
        &  4390.572  &  -0.53  &  80650.022  & $ 32.1 \pm 1.0 $ & $ -5.03 \pm 0.01 $ \\
        &  4433.988  &  -0.91  &  80650.022  & $ 22.6 \pm 1.1 $ & $ -4.79 \pm 0.02 $ \\
        &  4481.126  &   0.74  &  71490.190  & $ 232.6 \pm 1.7 $ & $ -5.08 \pm 0.01 $ \\
        &  4481.150  &  -0.56  &  71490.190  & \nodata\phn\phn & \nodata \\
        &  4481.325  &   0.59  &  71491.064  & \nodata\phn\phn & \nodata \\
        &  5401.521  &  -0.45  &  93799.630  & $ 13.2 \pm 1.8 $ & $ -5.14 \pm 0.05 $ \\
        &  5401.556  &  -0.34  &  93799.750  & \nodata\phn\phn & \nodata \\
        &  7877.054  &   0.39  &  80619.500  & $ 66.5 \pm 1.7 $ & $ -4.86 \pm 0.02 $ \\
        &  7896.042  &  -0.31  &  80650.022  & $ 95.1 \pm 2.9 $ & $ -5.03 \pm 0.02 $ \\
        &  7896.366  &   0.65  &  80650.022  & \nodata\phn\phn & \nodata \\
\altwo  &  4663.046  &  -0.28  &   85481.354  & $ 25.4 \pm 1.0 $ & $ -6.06 \pm 0.02 $ \\
        &  5593.300  &   0.41  &  106920.564  & $  3.4 \pm 2.2 $ & $ -6.46 \pm 0.26 $ \\
        &  6226.195  &   0.05  &  105427.522  & $  6.4 \pm 1.4 $ & $ -5.80 \pm 0.08 $ \\
        &  6231.750  &   0.40  &  105441.498  & $  3.8 \pm 1.3 $ & $ -6.54 \pm 0.15 $ \\
        &  6243.367  &   0.67  &  105470.928  & $ 20.2 \pm 1.2 $ & $ -5.89 \pm 0.03 $ \\
        &  7042.083  &   0.35  &   91274.503  & $ 21.8 \pm 1.4 $ & $ -6.11 \pm 0.04 $ \\
        &  7056.712  &   0.13  &   91274.503  & $  7.4 \pm 1.4 $ & $ -6.53 \pm 0.09 $ \\
        &  7063.682  &  -0.35  &   91274.503  & $ 10.8 \pm 1.5 $ & $ -5.85 \pm 0.06 $ \\
\sitwo  &  3853.665  &  -1.52  &  55309.353  & $ 69.6 \pm 3.2 $ & $ -4.99 \pm 0.03 $ \\
        &  3856.018  &  -0.56  &  55325.181  & $ 102.1 \pm 1.9 $ & $ -5.25 \pm 0.03 $ \\
        &  3862.595  &  -0.82  &  55309.353  & $ 95.9 \pm 2.5 $ & $ -5.15 \pm 0.03 $ \\
        &  4128.054  &   0.32  &  79338.502  & $ 82.2 \pm 1.2 $ & $ -5.20 \pm 0.02 $ \\
        &  4130.872  &  -0.82  &  79355.025  & $ 84.0 \pm 1.1 $ & $ -5.33 \pm 0.02 $ \\
        &  4130.894  &   0.48  &  79355.025  & \nodata\phn\phn & \nodata \\
        &  5041.024  &   0.29  &  81191.346  & $ 100.1 \pm 1.9 $ & $ -4.59 \pm 0.02 $ \\
        &  5055.984  &   0.59  &  81251.320  & $ 112.1 \pm 2.5 $ & $ -5.22 \pm 0.02 $ \\
        &  5056.317  &  -0.36  &  81251.320  & \nodata\phn\phn & \nodata \\
        &  5957.559  &  -0.30  &  81191.346  & $ 27.3 \pm 2.1 $ & $ -5.29 \pm 0.03 $ \\
        &  5978.930  &  0.00  &  81251.320  & $ 42.6 \pm 1.7 $ & $ -5.22 \pm 0.03 $ \\
        &  6371.371  &  0.00  &  65500.467  & $ 99.4 \pm 1.2 $ & $ -5.15 \pm 0.02 $ \\
\stwo  &  4162.665  &   0.83  &  128599.152  & $ 10.9 \pm 1.9 $ & $ -5.33 \pm 0.08 $ \\
       &  5212.620  &   0.66  &  121530.021  & $ 13.4 \pm 1.1 $ & $ -5.01 \pm 0.07 $ \\
       &  5428.655  &  -0.01  &  109560.686  & $ 11.4 \pm 1.7 $ & $ -5.03 \pm 0.08 $ \\
       &  5453.855  &   0.56  &  110268.606  & $ 21.4 \pm 1.5 $ & $ -5.05 \pm 0.05 $ \\
       &  5606.151  &   0.16  &  110766.562  & $  6.8 \pm 1.6 $ & $ -5.40 \pm 0.11 $ \\
       &  5639.977  &   0.33  &  113461.537  & $  9.0 \pm 1.0 $ & $ -5.28 \pm 0.06 $ \\
       &  5640.346  &   0.15  &  110508.706  & $  4.6 \pm 0.9 $ & $ -5.64 \pm 0.10 $ \\
\catwo  &  3736.902  &  -0.15  &  25414.399  & $ 22.8 \pm 3.3 $ & $-7.09 \pm 0.06$ \\
        &  3933.663  &   0.13  &  0.000  & $ 124.8 \pm 1.6  $ & $ -7.21 \pm 0.04 $ \\
        &  3968.469  &  -0.17  &  0.000  & $ 52.2 \pm 1.5 $ & $ -6.74 \pm 0.04 $\\
\titwo  &  3741.635  &  -0.11  &  12758.110  & $ 12.4 \pm 2.2 $ & $ -8.16 \pm 0.08 $ \\
        &  3759.296  &  0.20  &  4897.650  & $ 12.3 \pm 2.3 $ & $ -8.57 \pm 0.04 $ \\
        &  3761.323  &  0.10  &  4628.580  & $ 23.1 \pm 2.4 $ & $ -8.57 \pm 0.04 $ \\
\crtwo &  4242.364  &  -0.59  &  31219.350  & $ 14.2 \pm 1.5 $ & $ -7.90 \pm 0.04 $ \\
       &  4558.650  &  -0.66  &  32854.308  & $ 24.6 \pm 0.9 $ & $ -7.23 \pm 0.01 $ \\
       &  4588.199  &  -0.63  &  32836.680  & $ 21.9 \pm 1.0 $ & $ -7.33 \pm 0.02 $ \\
\fetwo &  3779.576  &  -3.78  &  20516.961  & $ 3.7 \pm 2.3 $ & $ -5.48 \pm 0.15 $ \\
       &  3781.509  &  -2.77  &  36252.917  & $ 5.4 \pm 2.2 $ & $ -5.30 \pm 0.15 $ \\
       &  3783.347  &  -3.16  &  18360.647  & $ 16.1 \pm 2.1 $ & $ -5.52 \pm 0.04 $ \\
       &  3845.180  &  -2.29  &  36126.386  & $ 11.6 \pm 2.2 $ & $ -5.45 \pm 0.06 $ \\
       &  4173.461  &  -2.18  &  20830.582  & $ 34.3 \pm 1.3 $ & $ -5.93 \pm 0.02 $ \\
       &  4177.692  &  -3.75  &  20516.961  & $ 9.3 \pm 1.2 $ & $ -5.12 \pm 0.03 $ \\
       &  4178.862  &  -2.48  &  20830.582  & $ 36.3 \pm 1.2 $ & $ -5.59 \pm 0.01 $ \\
       &  4303.176  &  -2.49  &  21812.055  & $ 35.7 \pm 1.1 $ & $ -5.52 \pm 0.01 $ \\
       &  4351.769  &  -2.10  &  21812.055  & $ 52.6 \pm 0.9 $ & $ -5.53 \pm 0.01 $ \\
       &  4555.893  &  -2.29  &  22810.356  & $ 43.0 \pm 0.9 $ & $ -5.51 \pm 0.01 $ \\
       &  4583.837  &  -2.02  &  22637.205  & $ 67.4 \pm 0.9 $ & $ -5.35 \pm 0.01 $ \\
       &  4629.339  &  -2.37  &  22637.205  & $ 45.2 \pm 0.8 $ & $ -5.41 \pm 0.01 $ \\
       &  5018.440  &  -1.22  &  23317.632  & $ 89.1 \pm 2.7 $ & $ -5.61 \pm 0.04 $ \\
       &  5169.033  &  -0.87  &  23317.632  & $ 112.2 \pm 2.0 $ & $ -5.45 \pm 0.03 $ \\
       &  5234.625  &  -2.05  &  25981.630  & $ 45.0 \pm 2.6 $ & $ -5.51 \pm 0.03 $ \\
       &  5276.002  &  -1.94  &  25805.329  & $ 68.1 \pm 1.5 $ & $ -5.16 \pm 0.02 $ \\
       &  5316.225  &  0.34  &  84035.139  & $ 89.0 \pm 2.5 $ & $ -5.33 \pm 0.02 $ \\
       &  5316.615  &  -1.85  &  25428.783  & \nodata & \nodata \\
       &  5316.784  &  -2.91  &  25981.630  & \nodata & \nodata \\
\nitwo &  3849.554  &  -1.88  &  32523.540  & $ 18.9 \pm 2.4 $ & $ -6.59 \pm 0.05 $ \\
       &  4067.031  &  -1.29  &  32499.529  & $ 16.7 \pm 2.0 $ & $ -7.25 \pm 0.06 $ \\
\enddata
\tablecomments{Abundances relative to hydrogen: \abund{X}.}
\end{deluxetable*}







\end{document}